\documentclass[aps,prb,twocolumn
,groupedaddress
,superscriptaddress
,amsfonts,
citeautoscript,a4paper]{revtex4-2}
\usepackage[T1]{fontenc}
\usepackage{amsmath}
\usepackage{amssymb}
\usepackage{graphicx}
\usepackage{babel}
\usepackage[normalem]{ulem}
\usepackage{color}
\usepackage{hyperref}
\hypersetup{colorlinks=true,allcolors=blue}
\usepackage{orcidlink}

\begin{document}
\title{Topology in a one-dimensional plasmonic crystal}

\author{D. A. Miranda\orcidlink{0000-0002-9784-3340}}
\affiliation{Centro de F\'{\i}sica (CF-UM-UP) and Departamento de F\'{\i}sica, Universidade do Minho, P-4710-057 Braga, Portugal}
\affiliation{POLIMA---Center for Polariton-driven Light--Matter Interactions, University of Southern Denmark, Campusvej 55, DK-5230 Odense M, Denmark}

\author{Y. V. Bludov\,\orcidlink{0000-0001-9648-1459}}
\affiliation{Centro de F\'{\i}sica (CF-UM-UP) and Departamento de F\'{\i}sica, Universidade do Minho, P-4710-057 Braga, Portugal}

\author{N. Asger Mortensen\,\orcidlink{0000-0001-7936-6264}}
\affiliation{POLIMA---Center for Polariton-driven Light--Matter Interactions, University of Southern Denmark, Campusvej 55, DK-5230 Odense M, Denmark}
\affiliation{Danish Institute for Advanced Study, University of Southern Denmark, Campusvej 55, DK-5230 Odense M, Denmark}

\author{N.~M.~R.~Peres\,\orcidlink{0000-0002-7928-8005}}

\affiliation{Centro de F\'{\i}sica (CF-UM-UP) and Departamento de F\'{\i}sica, Universidade do Minho, P-4710-057 Braga, Portugal}
\affiliation{POLIMA---Center for Polariton-driven Light--Matter Interactions, University of Southern Denmark, Campusvej 55, DK-5230 Odense M, Denmark}
\affiliation{International Iberian Nanotechnology Laboratory (INL), Av Mestre Jos\'e Veiga, 4715-330 Braga, Portugal}

\begin{abstract}
In this paper we study the topology of the bands of a plasmonic
crystal composed of graphene and of a metallic grating. Firstly, we derive
a Kronig--Penney type of equation for the plasmonic bands as function
of the Bloch wavevector and discuss the propagation of the surface
plasmon polaritons on the polaritonic crystal using a transfer-matrix approach considering a finite relaxation time. 
Second, we reformulate the problem as a tight-binding model
that resembles the Su--Schrieffer-Heeger (SSH) Hamiltonian, one difference being that the hopping amplitudes are, in this case, energy dependent.
In possession of the tight-binding equations it is a simple task to
determine the topology (value of the winding number) of the bands. This allows to determine the existense or absence of topological end modes in the system
Similarly to the SSH model, we
show that there is a tunable parameter that induces topological phase
transitions from trivial to non-trivial. In our case, it is the distance
$d$ between the graphene sheet and the metallic grating.  We note
that $d$ is a parameter that can be easily tuned experimentally simply
by controlling the thickness of the spacer between the grating and
the graphene sheet. It is then experimentally feasible to engineer
devices with the required topological properties.  
Finally, we propose a scattering experiment allowing the observation of the topological states.
\end{abstract}
\maketitle

\section{Introduction}

It is well known~\cite{Polman2005,Perenboom1981,Bashevoy2006} that
continuous metallic interfaces and metallic nanoparticles sustain
collective charge excitations, dubbed plasmon polaritons. In the case
of nanoparticles, this type of excitations are localized at their
external boundary~\cite{Mie1908,Fan2014}. For metallic films, the
charge excitations are also localized at the metal vacuum interface
but, contrary to metallic nanoparticles, they are free to travel along the
interface. This type of excitation, half matter, half light, is dubbed
surface plasmon polariton. In noble metals, such as silver and gold,
plasmon polaritons are sustained in the visible and near-IR spectral
range. For lower frequencies the plasmon polaritons become severely
damped due to Ohmic losses~\cite{Khurgin2015}. Often, devices based
on plasmon polaritons explore the highly confined electric fields
for applications such as sensing, optoelectronic applications, and
probing~\cite{Zeng2014,Yu2019,Gadelha2021}.

The severe aforementioned Ohmic losses call for other materials that
can sustain undamped plasmon polaritons. It has been shown that graphene
fills this gap by exhibiting both propagating and localized surface
plasmons in the mid-infrared (IR) spectrum, that can be tuned by gating or chemical doping~\cite{Jablan2009,Grigorenko2012,Ni2018,Fei2012}.
When encapsulated in hexagonal boron nitride, surface plasmon polaritons
in graphene show higher propagation lengths and stronger electromagnetic
field confinement~\cite{Woessner2015,Brar2013,Koppens2011}. This
property made graphene suitable for sensing and optoelectronic applications in the mid-IR spectral range~\cite{Rodrigo2015,Vasic2013,Li2014}.

\begin{figure}[b!]
\begin{centering}
\includegraphics[scale=0.4]{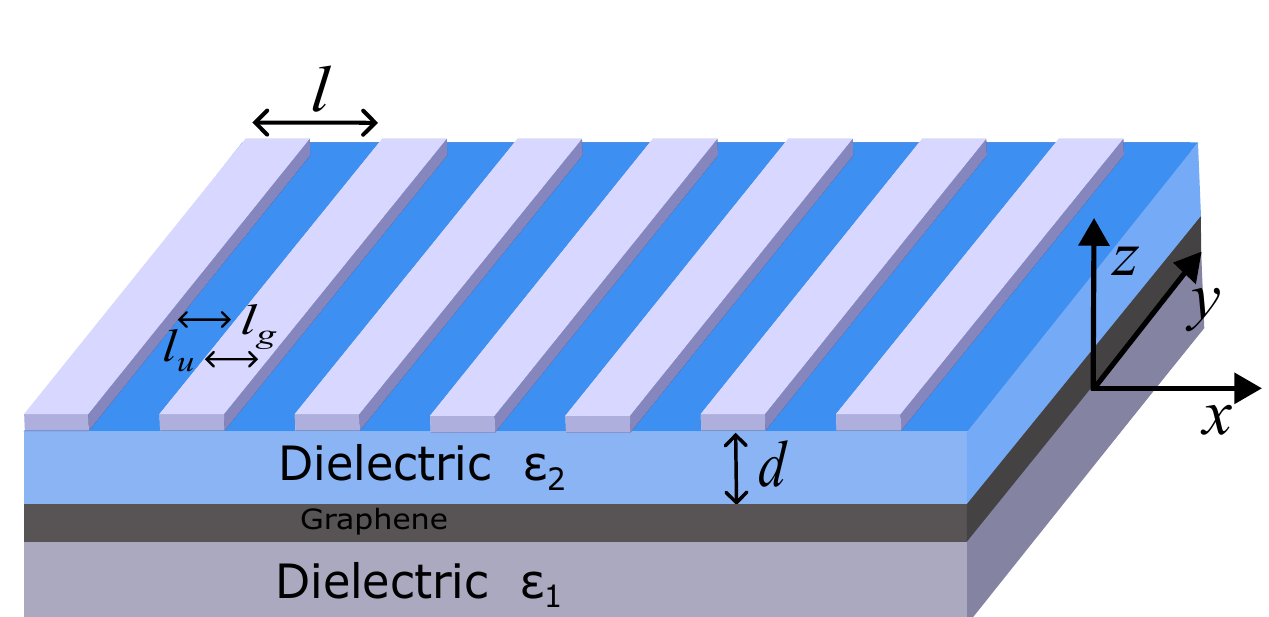}\caption{Representation of a plasmonic crystal periodic in the $x$ direction.
A graphene sheet is separated from a metal grating by a dielectric
of thickness $d$ and dielectric constant $\epsilon_{2}$. Bellow
the graphene sheet, another semi-infinite dielectric, with dielectric
constant $\epsilon_{1}$, acts as substrate. The metal grating can
be divided in two parts: a metallic one with length $l_{g}$, named
gated region, followed by a stripe where the metal is absent, and
with length $l_{u}$, named ungated region. The unit cell of the plasmonic
crystal has length $l=l_{g}+l_{u}$.}
\par\end{centering}
\end{figure}

It has been shown that plasmons in noble metal based structures can,
in certain circumstances, sustain topological plasmons, as is the case
of structured silver films where topological plasmonic vortices have
been observed~\cite{Dai2020}. Topology in structured materials is
actually a quite general feature of matter, exhibited in systems as
diverse as photonic and phononic devices~\cite{Gong2021,Kim,mech_SSH1,mech_SSH2,mech_SSH3,mech_SSH4,Shi2021},
and cold atom lattices~\cite{Zhang2018,Wintersperger2020}, and being
rather ubiquitous in condensed matter systems~\cite{Konig2007,Fu2007,Chen2009,SSH_original},
where the crystalline structure plays an essential role in determining
the presence or absence of topological energy bands. 

Graphene itself is not a topological material (although it was shown
that a hexagonal array of metallic nanoparticles can be~\cite{Zhang2023}).
Therefore, the question arises if structured graphene can exhibit
topological properties. In particular, if plasmon polaritons in this
material can present a polaritonic band structure with topological
bands. A preliminary and fully numerical study suggests that this
is indeed the case~\cite{Dafei2017}. Graphene can be structured in
several different ways. Two possibilities are patterning a periodic
array of graphene ribbons~\cite{Ju2011,Nikitin2012,Strait2013,Bo2015}
or coupling a continuous graphene sheet with a metallic grating~\cite{Dias2017,Bylinkin2019,Rappoport2021,Guo23}.
The advantage of the second approach is twofold. First, no additional
scattering channels are introduced in the system as opposed to the
patterned case, which leads to additional scattering of the surface
plasmons at the rough edges of the ribbons. Secondly, the micro-fabrication
process is much facilitated. We, therefore, focus our attention on
the second system. 

It is well known~\cite{Bylinkin2019,Rappoport2021,GraphenePlasmons}
that graphene surface plasmons behave considerably different when
in the vicinity of a metal substrate, in comparison to usual graphene
plasmons when graphene is deposited on a dielectric substrate alone.
In the latter case the plasmon are dubbed graphene plasmons (or ungated
plasmons), whereas in the first case they are dubbed screened graphene
plasmons (occasionally also refereed to as acoustic plasmons, gated plasmons, or even image polaritons~\cite{image-polariton}). This distinction
is made because, in the presence of a metal, the electric field is
screened, leading to a softening of the plasmon dispersion, that becomes
linear with the wave vector~\cite{GraphenePlasmons}, as opposed to
the usual graphene plasmons which disperse with the square root of
the wave vector~\cite{GraphenePlasmons}. Thus, graphene in the vicinity
of a metallic grating is a hybrid system with alternating regions
of gated and ungated plasmons. We call this system a plasmonic crystal.
The unit cell is then composed of a region of gated plasmons (acoustic
graphene plasmons with $\omega \propto q_g$, where $q_g$ is a wavenumber) followed by a region of ungated ones (graphene
plasmons with $\omega \propto \sqrt{q_u}$, where $q_u$ is a wavenumber). The fact that the dispersion of the two types of plasmons
is different, as noted before, has the consequence that both types
of plasmons have different wave vector for the plasmon polaritons
of a give frequency. From the point of view of a theoretical description,
the envisioned system is much like a Kronig--Penney model~\cite{KP_original}
for electrons moving in periodic arrangement of electrostatic potentials.
Indeed, when we describe the system in terms of electrostatic potentials
and electric current densities, the transcendental equation leading
to the polaritonic bands is formally identical to that of the Kronig--Penney model. However, the Kronig--Penney model formalism is not simple to analyze due to its infinite number of bands and lack of fully analytical expression for the Boch functions. Also, from the point
of view of topology, the original formulation of the model lacks the
form of an electronic tight-binding Hamiltonian, for which well known
tools exist for describing the system's topology. One such electronic
model is the one-dimensional Su--Schrieffer--Heeger (SSH) model~\cite{2016short,SSH_original}.
Electrons described by the SSH model are known to show topological
behavior when certain conditions are met for the hopping parameters.
The question now arises if it is possible to describe our system, a
periodic array of gated and ungated graphene plasmons, with the SSH
model formalism. Our analysis below shows that this is indeed the
case. With this mapping in hands, the description of topology becomes
much simpler than in the original Kronig--Penney formulation.
Finally,  we note that our results can also be applied to other system beyond
graphene \cite{Dyer2013,Shuvaev2022}.

This paper is organized as follows. In Sec.~\ref{sec:Crystal_bands_trans}
we introduce the polaritonic crystal, as realized in Ref.~\onlinecite{Bylinkin2019}
alongside the mathematical formalism. Thereafter, we derive the plasmonic
energy dispersion and the crystal propagation properties. In Sec.~\ref{sec:SSH-TB} the SSH tight-binding representation of the plasmonic crystal is introduced.
There, we investigate how the distance between the metal grating and
graphene affects the energy-dependent tight-binding coefficients and
the topology associated with the plasmonic bands. Additionally, we
derive an analytical expression for the particular distance that results in
the energy gaps closing, as well as an analytical approximation for
the bands. Finally, we offer our conclusions and detail important
calculations in the Appendixes. 

\section{Plasmons in  graphene in the presence of a metallic grating\label{sec:Crystal_bands_trans}}

In this system, experimentally studied in Ref.~\onlinecite{Bylinkin2019}, we
have a graphene sheet sandwiched between two dielectrics, characterized
by dielectric constants $\epsilon_{1}$ and $\epsilon_{2}$. On the
upper-most dielectric, at a distance $d$ from the graphene sheet,
there is a metal grating. The grating is a one-dimensional (1D) periodic structure composed
of two alternating portions: a metallic section of length $l_{g}$
(denominated gated region) and an empty portion of length $l_{u}$(named
ungated region), such that the unit cell of length $l=l_{u}+l_{g}$
can be defined. The presence of a conductor significantly alters the
properties of plasmons on graphene. For instance, the wavenumbers
of the gated ($q_{g})$ and ungated $(q_{u})$ regions differ, being given by~\cite{GraphenePlasmons}
\begin{subequations}
\begin{equation}
q_{g}(\omega,d)=\omega\sqrt{\frac{\hbar\epsilon_{2}}{4E_{F}c\alpha d}},\label{eq:q_gated}
\end{equation}
\begin{equation}
q_{u}(\omega)=\frac{\hbar(\epsilon_{1}+\epsilon_{2})\omega^{2}}{4E_{F}c\alpha},\label{eq:q_ungated}
\end{equation}
\end{subequations}
respectively, where $E_{F}$ is the Fermi energy of graphene (considered
to be equal in both regions), $\alpha$ is the fine-structure constant,
$c$ is the speed of light in vacuum, and $\omega$ is the plasmon
frequency. We note that Eqs.~(\ref{eq:q_gated}) and (\ref{eq:q_ungated})
are analytic approximations to the exact dispersion, which can be
known only numerically. However, the main approximation implicit in these two equations is  valid as it corresponds to the non-retarded limit. This limit is accurate in plasmonics due to the large wavenumbers of the fields compared with $\omega/c$. We also note that $q_{g/u}(\omega,d)$ is always real because we are considering only normal incidence \cite{Rejaei_2015} and neglecting damping.  In the case of oblique incidence $q_{g/u}(\omega,d)$ can be imaginary (evanescent waves) due to the finite value of $k_y$ \cite{Alymov2023}, even in the absence of damping.  Considering finite $k_y$ complicates the forthcoming analysis and adds no new physics to the problem.
In summary, to obtain Eqs.~(\ref{eq:q_gated}) and (\ref{eq:q_ungated}),  it is necessary to assume
that (i) the frequency regime is such that the conductivity of graphene
is dominated by its Drude conductivity $\sigma(\omega)=\frac{4iE_{F}}{(\hbar\omega+i\Gamma)}\alpha\epsilon_{0}c$,
(ii) absorption is zero ($\Gamma=0)$, and, most importantly, (iii) we can approximate
the non retarded limit, $q_{u/g}\gg\frac{\sqrt{\epsilon_{j}}\omega}{c}$,
where $j=1,2$ is the index labeling the dielectrics. Also, we consider
values of $d$ such that $q_{g}d\ll1$, which reflects the experimental situation ~\cite{Iranzo2018}. The calculations leading
to Eqs.~(\ref{eq:q_gated}) and (\ref{eq:q_ungated}) are detailed in
Appendix~A. 

Our goal in this section is to use the transfer-matrix formalism to
understand the propagation of graphene surface plasmons and to obtain
the energy dispersion, associating the Bloch momentum $q$, arising
from the periodic arrangement of the metal grating, with the energy
$\hbar\omega$. 

\subsection{Fields, boundary conditions and transfer matrices}

\begin{figure}
	\centering{}\includegraphics[scale=0.45]{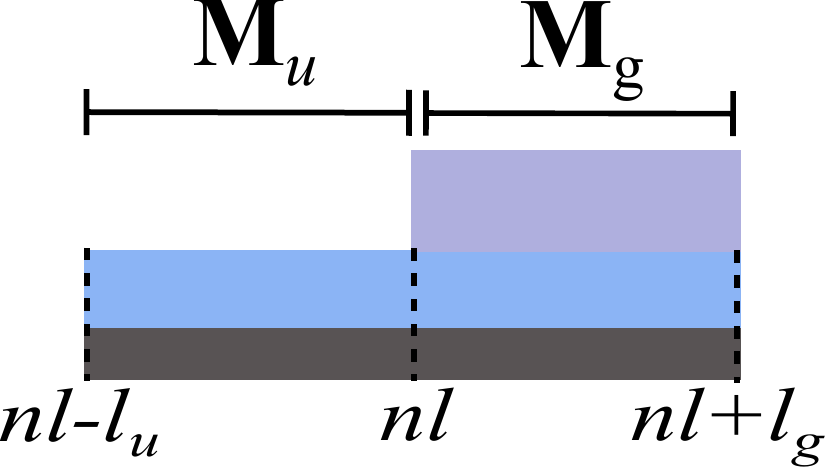}\caption{\label{fig:unit_cell_transfer_matrix}Scheme of an unit cell beginning
		at $x=nl-l_{u}$ and finishing at $x=nl+l_{g}$ from a front view. The gated region
		is comprised between $nl<x<nl+l_{g}$ and the ungated region is comprised
		between $nl-l_{u}<x<nl$. The transfer matrix relating the wave amplitudes
		to the left of the gated region with the wave amplitudes to the right
		of the gated region is ${\bf M}_{g}$. The transfer matrix relating
		the wave amplitudes to the left of the gated region with the wave
		amplitudes to the right of the ungated region is ${\bf M}_{u}$. The
		unit cell transfer matrix is then defined as ${\bf M}={\bf M}_{u}{\bf M}_{g}$.}
\end{figure}

We begin defining the expressions of the electrostatic potential
describing propagating plasmons in the ungated and gated regions, in the $n^{th}$ unit cell 
\begin{subequations}    
\begin{equation}
\Phi_{u}^{n}(x)=\phi_{u+}^{n}e^{iq_{u}(x+l_u-nl)}+\phi_{u-}^{n}e^{-iq_{u}(x+l_u-nl)},\label{eq:phi_u and phi_g_1}
\end{equation}
\begin{equation}    
\Phi_{g}^{n}(x)=\phi_{g+}^{n}e^{iq_{g}(x-nl)}+\phi_{g-}^{n}e^{-iq_{g}(x-nl)},\label{eq:phi_u and phi_g_2}
\end{equation}
\end{subequations}
where the $+(-)$ subscript correspond to the wave component traveling
to the right (left) and the superscript $n$ labels the unit cell the potential is being evaluated at. We can think of the gated regions as rectangular
barriers of length $l_{g}$ that begin at $x=nl$ and end at $x=nl+l_{g}$, and of ungated regions, that begin at $nl-l_u$ and end at $nl$, where $n\in\mathbb{Z}$ (for details see Fig.~\ref{fig:unit_cell_transfer_matrix}). We can then make use of appropriate boundary conditions to relate the electrostatic potential between gated and ungated regions. At the boundaries between regions we impose: (i) the continuity of the electrostatic potential $\Phi(x)$ and (ii) the continuity of the current density $-\sigma\frac{\partial\Phi(x)}{\partial x}$ = $j(x)$, where $\sigma$ is the graphene conductivity (Drude conductivity) defined at the beginning
of the section. At $x=-l_{u}+nl$, we relate the gated potential at unit cell $n-1$ to the ungated potential at unit cell $n$. The same must be done to the current density:
\begin{subequations}    
\begin{equation}
\Phi_{g}^{n-1}(nl-l_{u}) = \Phi_{u}^{n}(nl-l_{u})
,\label{eq:BCx=00003Dlg1}
\end{equation}
\begin{eqnarray}
\sigma_{g}\frac{\partial\Phi_{g}^{n-1}(nl-l_{u})}{\partial x}=\sigma_{u}\frac{\partial\Phi_{u}^{n}(nl-l_{u})}{\partial x},\label{eq:BCx=00003Dlg2}
\end{eqnarray}
\end{subequations}
while at $x=nl$ we relate ungated and gated potentials and current densities of the same unit cell $n$:
\begin{subequations}    
\begin{equation}
\Phi_{u}^{n}(nl) = \Phi_{g}^{n}(nl),\label{eq:BCx=00003D0_1}
\end{equation}
\begin{eqnarray}
\sigma_{g}\frac{\partial\Phi_{g}^{n}(nl)}{\partial x}=\sigma_{u}\frac{\partial\Phi_{u}^{n}(nl)}{\partial x}.\label{eq:BCx=00003D0_2}
\end{eqnarray}
\end{subequations}

We then define the gated and ungated region transfer matrices, according
to Fig.~\ref{fig:unit_cell_transfer_matrix}. These matrices relate the potential amplitudes $\phi$ at the beginning of the region with the potential at the end of the region:
\begin{subequations}    
\begin{eqnarray}
\left(\begin{array}{c}
\phi_{g+}^{n}\\
\\
\phi_{g-}^{n}
\end{array}\right)={\bf M}_{u}\left(\begin{array}{c}
\phi_{u+}^{n}\\
\\
\phi_{u-}^{n}
\end{array}\right)\\
\left(\begin{array}{c}
\phi_{u+}^{n}\\
\\
\phi_{u-}^{n}
\end{array}\right)={\bf M}_{g}\left(\begin{array}{c}
\phi_{g+}^{n-1}\\
\\
\phi_{g-}^{n-1}
\end{array}\right)
\end{eqnarray}
\end{subequations}
So, the transfer matrix ${\bf M}_{u}$ relates the gated and ungated potential amplitudes in the same unit cell, while ${\bf M}_{g}$ relates the ungated amplitudes of unit cell $n$ with the gated potential amplitudes of the unit cell $n-1$. The transfer matrices ${\bf M}_{g}$ can be obtained from Eqs.~\eqref{eq:BCx=00003Dlg1} and \eqref{eq:BCx=00003Dlg2} by writing the potentials explicitly as in Eqs.~\eqref{eq:phi_u and phi_g_1} and \eqref{eq:phi_u and phi_g_2}. In a similar manner, using Eqs.~\eqref{eq:BCx=00003D0_1} and \eqref{eq:BCx=00003D0_2} alongside the explicit form of the potentials yields ${\bf M}_{u}$. They will have the common form:
\begin{equation}
{\bf M}_{\alpha}=\tfrac{1}{2}\left(\begin{array}{cc}
\left(1+\frac{\sigma_{\alpha}q_{\alpha}}{\sigma_{\beta}q_{\beta}}\right)e^{iq_{\alpha}l_{\alpha}} & \left(1-\frac{\sigma_{\alpha}q_{\alpha}}{\sigma_{\beta}q_{\beta}}\right)e^{-iq_{\alpha}l_{\alpha}}\\\\
\vspace{-7mm}\\
\left(1-\frac{\sigma_{\alpha}q_{\alpha}}{\sigma_{\beta}q_{\beta}}\right)e^{iq_{\alpha}l_{\alpha}} & \left(1+\frac{\sigma_{\alpha}q_{\alpha}}{\sigma_{\beta}q_{\beta}}\right)e^{-iq_{\alpha}l_{\alpha}}
\end{array}\right).
\label{eq:Mg}
\end{equation}
where $\alpha, \beta$ are general region indexes. If $\alpha=q, \beta=u$ and vice-versa. 
\subsection{Dispersion and energy bands}
\begin{figure}
	\centering{}\includegraphics[scale=0.4]{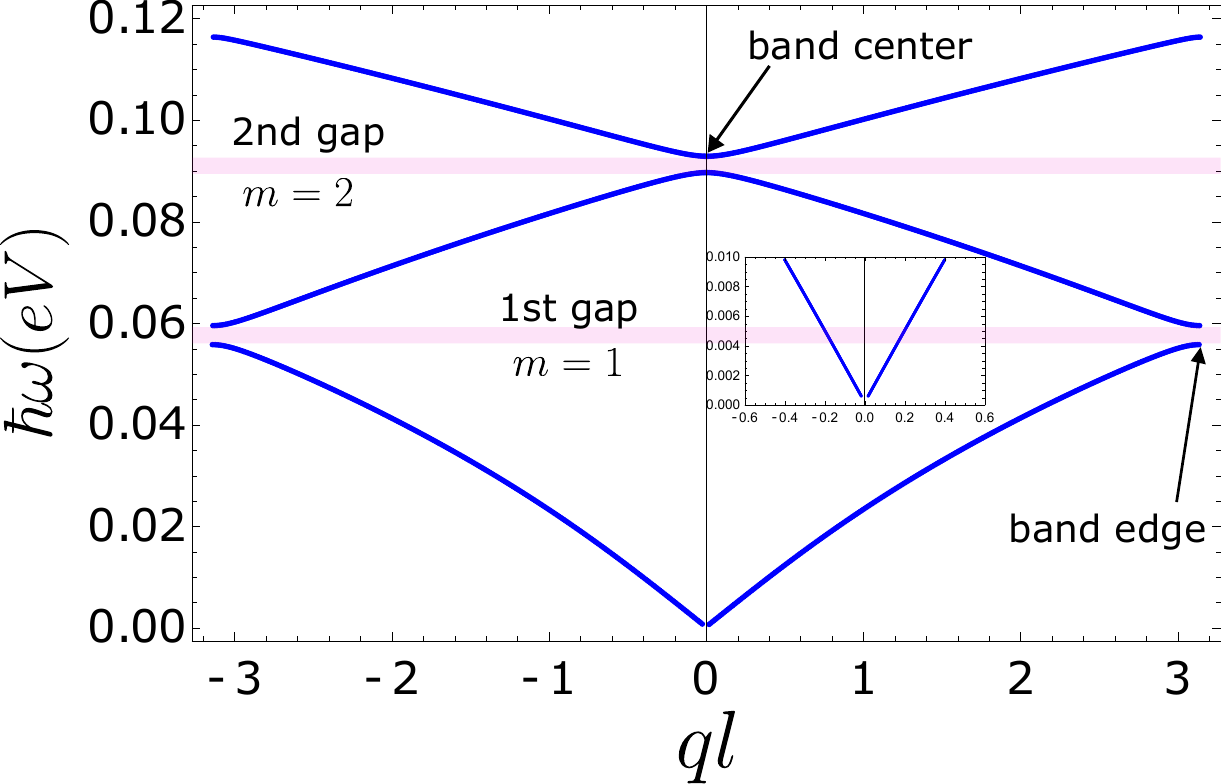}\caption{\label{fig:KP_dispersion}First three energy bands as a function
		of the Bloch wave number $ql$. The first two energy band gaps are highlighted in red, with the respective gap index $m$. The parameters used in this numerical calculation were $l=500$\,nm, $l_{g}=240$\,nm, $l_{u}=260$\,nm, $d=100$\,nm, and $E_{F}=0.45$\,eV. The dielectrics encapsulating the graphene sheet were
		considered air ($\epsilon_{1}=1)$ and h-BN ($\epsilon_{2}=3.5)$,
		such that the metal grating is on top of h-BN. The inset in the figure
		shows the dispersion relation closer to $ql=0$ where it is possible
		to see that, for very small $q$, the dispersion is linear. }
\end{figure}

Coefficients of the electrostatic potential across one period of the structure, at $x=(n-1)l$ and $x=nl$, can be related by the unit cell transfer matrix ${\bf M}={\bf M}_{u}{\bf M}_{g}$. Consequently,
we must then have:
\begin{equation}
\left(\begin{array}{c}
\phi_{g+}^{n}\\
\\
\phi_{g-}^{n}
\end{array}\right)={\bf M}\left(\begin{array}{c}
\phi_{g+}^{n-1}\\
\\
\phi_{g-}^{n-1}
\end{array}\right).
\label{eq:M}
\end{equation}
The matrix ${\bf M}={\bf M}_{u}{\bf M}_{g}$ is unimodular, i.e., ${\rm det}\left({\bf M}\right)=1$.
The Bloch momentum $q$ is introduced by invoking Bloch theorem:
\begin{equation}
\left(\begin{array}{c}
\phi_{g+}^{n}\\
\\
\phi_{g-}^{n}
\end{array}\right)=e^{iqnl}\left(\begin{array}{c}
\phi_{g+}^{0}\\
\\
\phi_{g-}^{0}
\end{array}\right).
\label{eq:Bloch}
\end{equation}
Non-trivial solutions of the system of Eqs.~\eqref{eq:M} and \eqref{eq:Bloch} for the coefficients $\phi_{g+}^{0},\phi_{g-}^{0}$ are obtained if the determinant vanishes, that is
\begin{equation}
{\rm det}\left({\bf M}-e^{iql}{\bf I}\right)=0,
\label{eq:dr-prelim}
\end{equation}
where ${\bf I}$ is the identity matrix.
After taking into account the property of unimodular matrices, Eq.~\eqref{eq:dr-prelim} can be represented as~\cite{WaveProp}
\begin{equation}
2\cos(ql)={\rm Tr}({\bf M}),
\label{eq:dr}
\end{equation}
where ${\rm Tr}$ stands to the trace operation.
This condition yields a Kronig--Penney-like dispersion relation for the graphene plasmons which, after taking the explicit form of matrices ${\bf M}_u$ and ${\bf M}_g$ into account, can be written in the form
\begin{equation}
{\rm cos}(ql)={\rm cos}(q_{u}l_{u}){\rm cos}(q_{g}l_{g})-Z{\rm sin}(q_{u}l_{u}){\rm sin}(q_{g}l_{g}),\label{eq:K-P plasmon dispersion}
\end{equation}  where $Z=\left[\frac{\sigma_{u}q_{u}}{2\sigma_{g}q_{g}}+\frac{\sigma_{g}q_{g}}{2\sigma_{u}q_{u}}\right]$.
The numerical solution of the dispersion relation~\eqref{eq:K-P plasmon dispersion} is shown in Fig.~\ref{fig:KP_dispersion}. The periodicity of the structure results in a spectrum that consists in energy bands (where propagation of waves is allowed) and forbidden gaps (where the propagation is exponentially suppressed). Notice that in the low-frequency region the dispersion is linear (as shown in inset of Fig.~\ref{fig:KP_dispersion}). In this regime, we can expand Eq.~\eqref{eq:K-P plasmon dispersion} into a series with respect to the frequency $\omega$, obtaining the following approximation 
\begin{equation}
\left|q\right|\approx\omega\frac{l_g}{l}\sqrt{\frac{\hbar\epsilon_{2}}{4E_{F}c\alpha d}\left(1+\frac{\sigma_gl_u}{\sigma_ul_g}\right)}.
\end{equation} 

\subsection{Propagation of graphene plasmons}

\begin{figure}
\centering{}\includegraphics[scale=0.22]{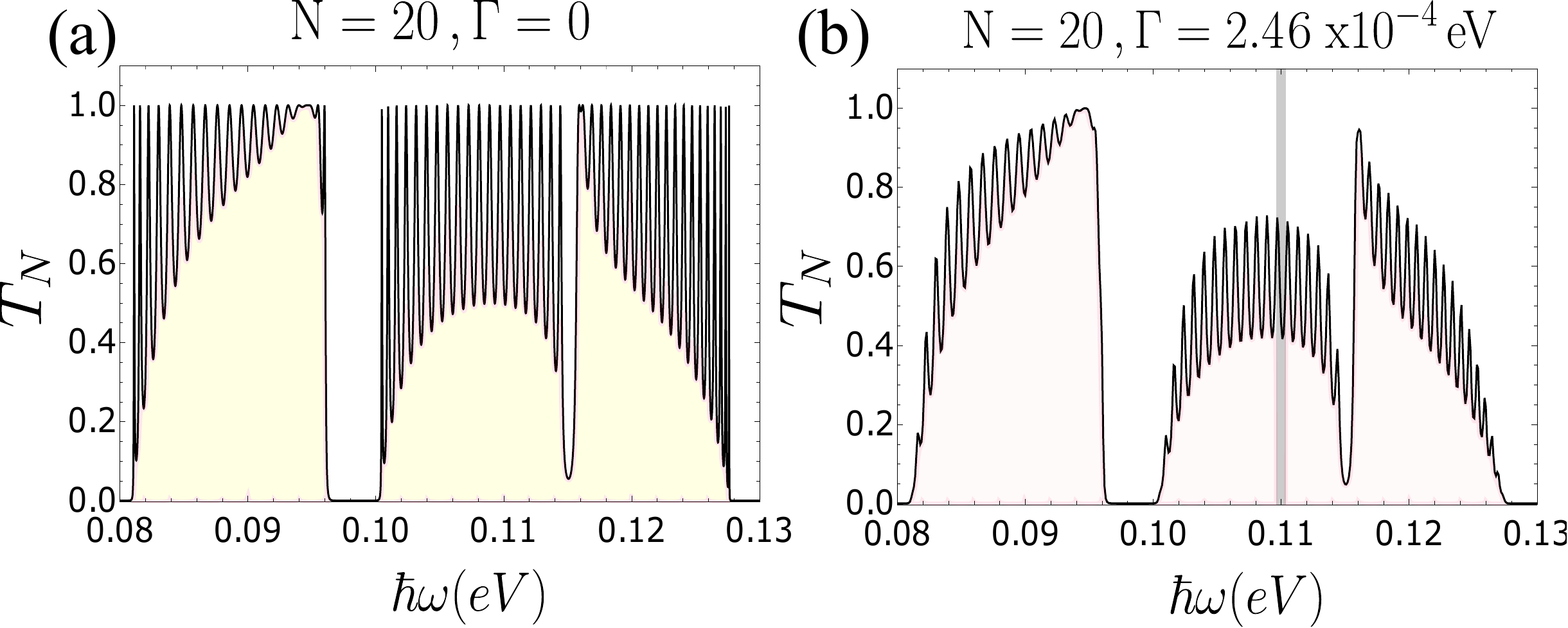}
\caption{System's transmission coefficient
$T_{N}$ as a function of the energy $\hbar\omega$ for $20$ unit cells and $\mathbf{(a)}$ $\Gamma=0$; $\mathbf{(b)}$ $\Gamma=2.46\rm{x}10^{-4}$ eV. 
The gray vertical line in $\mathbf{(b)}$ represents the frequency of the experimental determination of the plasmon damping at $T=60$ K .
The parameters used in this figure were $l=500$\,nm, $l_{g}=240$\,nm, $l_{u}=260$\,nm, $d=100$\,nm and $E_{F}=0.45$\,eV. The dielectrics
encapsulating the graphene sheet were considered h-BN ($\epsilon_{1}=\epsilon_{2}=3.5)$.  Similar results were found for graphene on Bragg gratings \cite{Peres_Bragg_2019}.}
\label{fig:Transport_vs_energy}
\end{figure}

With both ${\bf M}_{u}$ and ${\bf M}_{g}$, we have all necessary
tools to describe the propagation of the graphene plasmon polaritons
through the system. The transmission coefficient $T_{N}$ for a system with $N$ identical unit cells is given by~\cite{WaveProp}
\begin{equation}
T_{N}=\frac{1}{1+|{\bf M}_{12}^{N}|^{2}}
\end{equation}
where ${\bf M}^{N}$ is the transfer matrix of the system with $N$
unit cells, which is the product of $N$ matrices ${\bf M}$. The
element ${\bf M}_{12}^{N}$ is given by Chebyshev's identity~\cite{WaveProp}
as:
\begin{equation}
{\bf M}_{12}^{N}={\bf M}_{12}\frac{{\rm sin}(Nql)}{{\rm sin}(ql)}.
\end{equation}
The quantity $ql$ is a phase related to the Bloch momentum and obeys Eq.~\eqref{eq:dr}. Thus, ${\bf M}_{12}^{N}$
is easily obtainable after multiplying ${\bf M}_{u},{\bf M}_{g}$
and the computation of $T_{N}$ yields:

\begin{align}
T_{N} & =\frac{1}{1+\frac{{\rm sin^{2}}(Nql)}{{\rm sin^{2}}(ql)}|{\bf M}_{12}|^{2}}\\
 & =\frac{1}
 {1+\frac{{\rm sin^{2}}(Nql)}{{\rm sin^{2}}(ql)}\left|\frac{{\rm sin}(q_{u}l_{u})\left(\sigma_{g}^2q^2_{g}-\sigma_{u}^2q_{u}^2\right)}{2\sigma_{g}q_{g}\sigma_{u}q_{u}}\right|^{2}}.\nonumber 
\end{align}
It was reported in Ref. \cite{Ni2018} that h-BN encapsulated graphene supports plasmons with  propagation lengths of around $10\rm{\mu m}$ for an excitation energy of $0.11$ eV,  at cryogenic temperatures of around $T=60$ K. It was found that, at this energy and temperature, the scattering rate $\Gamma$ is around $2$ $\rm{cm}^{-1}$ or  $2.46\rm{x}10^{-4}$ eV. To verify if our plasmonic crystal also supports propagating plasmons in these conditions, we calculate the transmission
coefficient $T_{N}$ as a function of the energy $\hbar\omega$ for $N=20$ unit cells (which in our case corresponds to $10\rm{\mu m}$) for zero and finite scattering rate $\Gamma = 2.46\rm{x}10^{-4}$ eV. Both dielectrics were considered to be h-BN ($\epsilon$ = 3.5), such that Fig.~\ref{fig:Transport_vs_energy} (b) has the same parameters of the experiment in Ref. \cite{Ni2018}. We find that our system indeed supports plasmons with energy $0.11$ eV [gray vertical line in Fig.~\ref{fig:Transport_vs_energy} (b)], for which $T_{N}\approx0.7$. Moreover, we can identify three transport regimes in our plasmonic crystal: (i) zero transmittance zones, which correspond to gaps in the spectrum of the infinite plasmonic crystal, for which the condition $2<{\rm Tr}({\bf M})$ is valid. For these frequencies the periodic structure serves as a plasmonic Bragg mirror; (ii) $0 < T_{N} < 1$, corresponding to frequency ranges of allowed bands in the spectrum, for which transmittance exhibits periodic oscillations; (iii) $T_{N} = 1$, where the system becomes fully transparent. This occurs for frequencies corresponding to Bloch wavevectors $ql=n\pi/N$, where $1\le n \le N-1$. Naturally, this rarely occurs if $\Gamma\neq0$. It is also worth noting that, in reality, $\Gamma$ is a function of the frequency. The value $\Gamma = 2.46\rm{x}10^{-4}$ eV corresponds to the excitation energy $0.11$ eV, according to experiment \cite{Ni2018}. Thus, the values of $T_{N}$ depicted in Fig.~\ref{fig:Transport_vs_energy} (b) for other energies are approximation.

\section{Plasmonic SSH tight-binding model\label{sec:SSH-TB}}

\begin{figure}

\begin{centering}
\includegraphics[scale=0.17]{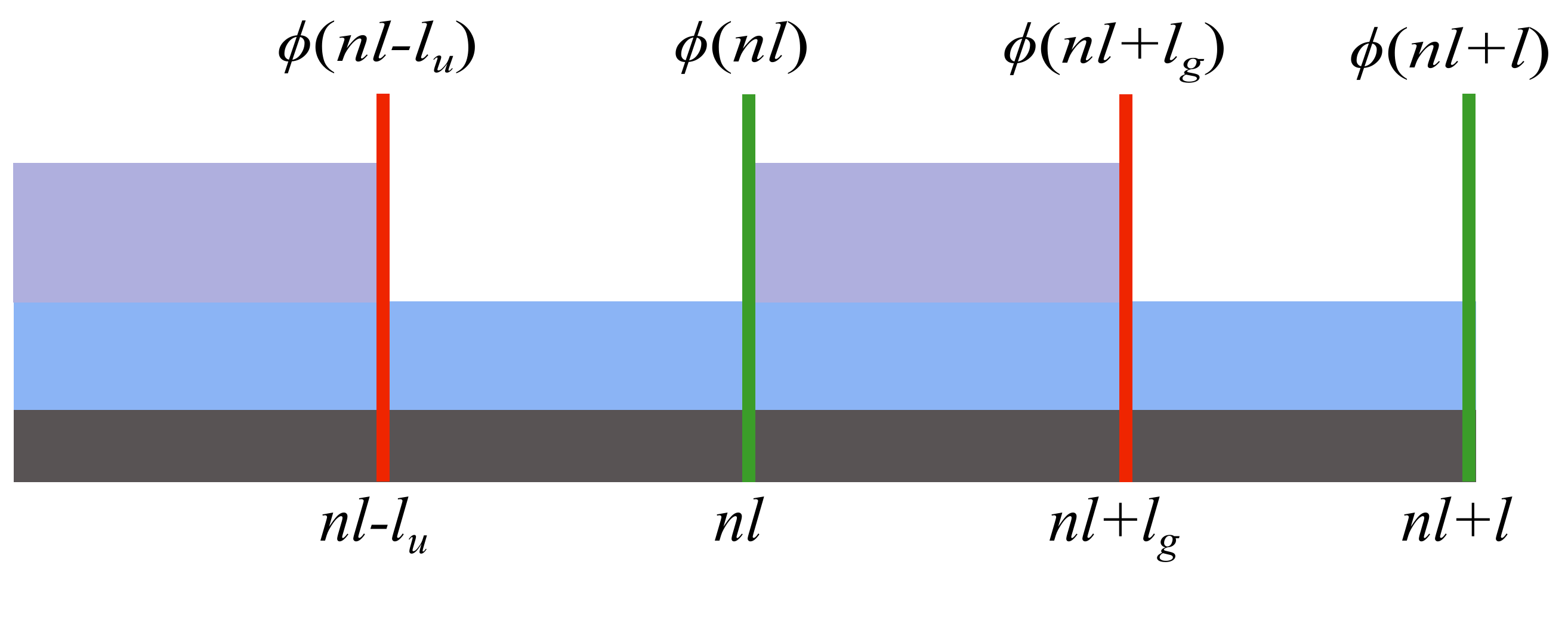}
\par\end{centering}
\centering{}\caption{\label{fig: two_sublattices_plasmonicSSH}Scheme of the SSH representation
of the plasmonic crystal. The two sub-lattices are depicted with different
colors: the red one has lattice sites at $x=nl+lg$ and the green
one at $x=nl$, where $n\in\mathbb{Z}$. Since these lattice sites
are defined at boundaries between gated and ungated regions, the sub-indexes
$u,g$ are irrelevant.}
\end{figure}

The goal in this section is to represent the electrostatic potential
describing the plasmons in the tight-binding formalism. This representation is rather useful for the discussion of the topological description of the bands. The symmetry of the problem (absence of inversion symmetry and two segments composing one unit cell) invites us to make an analogy with a system that possesses sub-lattice symmetry, like the SSH model.
To do this, we need to define the two sub-lattices: the first one
is defined by the points at $x=nl+lg$, depicted in red in Fig.~\ref{fig: two_sublattices_plasmonicSSH} and is associated with the
end of gated regions. The other sub-lattice is defined by points at
$x=nl$, depicted in green, and is associated with the end of ungated
regions. To derive the tight-binding representation we then need to
find expressions for the fields $\phi_{u}(x)$, $\phi_{g}(x)$, $\sigma_{u}\frac{\partial\phi_{u}(x)}{\partial x}$, and $\sigma_{g}\frac{\partial\phi_{g}(x)}{\partial x}$
at each lattice site as a function of the previous and the next sites
using the transfer-matrix formalism. For example, we want to obtain
$C\phi(nl+l_{g})=A_g\phi(nl)+A_u\phi(nl+l)$ and $C\phi(nl)=A_g\phi(nl+l_{g})+A_u\phi(nl-l_{u})$,
where $A$, $B$, and $C$ are the tight-binding coefficients to be determined.

\subsection{Energy-dependent hopping amplitudes}

The detailed derivation of the tight-binding equations can be found
in Appendix~B, where we show how to write the following equations
relating lattice sites from two different sub-lattices

\begin{widetext}
\begin{subequations}    
\begin{equation}
\phi(nl+l_{g})\frac{\sigma_{g}q_{g}}{{\rm sin}(q_{g}l_{g})}+\phi(nl-lu)\frac{\sigma_{u}q_{u}}{{\rm sin}(q_{u}l_{u})}=\phi(nl)\left[\frac{\sigma_{u}q_{u}{\rm cos}(q_{u}l_{u})}{{\rm sin}(q_{u}l_{u})}+\frac{\sigma_{g}q_{g}{\rm cos}(q_{g}l_{g})}{{\rm sin}(q_{g}l_{g})}\right]\label{eq:TB_eq_1}
\end{equation}
\begin{equation}
\phi(nl)\frac{\sigma_{g}q_{g}}{{\rm sin}(q_{g}l_{g})}+\phi(nl+l)\frac{\sigma_{u}q_{u}}{{\rm sin}(q_{u}l_{u})}=\phi(nl+l_{g})\left[\frac{\sigma_{u}q_{u}{\rm cos}(q_{u}l_{u})}{{\rm sin}(q_{u}l_{u})}+\frac{\sigma_{g}q_{g}{\rm cos}(q_{g}l_{g})}{{\rm sin}(q_{g}l_{g})}\right]\label{eq:TB_eq_2}
\end{equation}
\end{subequations}
\end{widetext} 
These equations relate the electrostatic potential in one "sub-lattice" with the electrostatic potential at nearest-neighbor
lattice points, contained in the other sub-lattice. The functions
multiplying the electrostatic potentials play the role of hopping
amplitudes in electronic tight-binding equations. To make Eqs.~(\ref{eq:TB_eq_1}) and (\ref{eq:TB_eq_2}) look more similar to the SSH tight-binding Hamiltonian,
we introduce
\begin{subequations}    
\begin{eqnarray}
    A_\alpha&=&\frac{\sigma_{\alpha}q_{\alpha}}{{\rm sin}(q_{\alpha}l_{\alpha})}, \alpha = \{u,g\}\\
C&=&\frac{\sigma_{u}q_{u}{\rm cos}(q_{u}l_{u})}{{\rm sin}(q_{u}l_{u})}+\frac{\sigma_{g}q_{g}{\rm cos}(q_{g}l_{g})}{{\rm sin}(q_{g}l_{g})}\\
&=&A_{u}{\rm cos}(q_{u}l_{u})+ A_{g}{\rm cos}(q_{g}l_{g})\\
\psi_{n}&=&\phi(nl)\\
\xi_{n}&=&\phi(nl+l_{g})
\end{eqnarray}
\label{eq:definition_coefficients_ABC}
\end{subequations}
which conveniently allows us to rewrite Eqs.~(\ref{eq:TB_eq_1}) and (\ref{eq:TB_eq_2}) as:
\begin{subequations}    
\begin{eqnarray}
C\psi_{n}=A_{g}\xi_{n}+A_{u}\xi_{n-1}\\
C\xi_{n}=A_{g}\psi_{n}+A_{u}\psi_{n+1}
\end{eqnarray}
\label{eq:plasmonic_SSH equations}
\end{subequations}
These coefficients are energy dependent through $q_{u}$, $q_{g}$, $\sigma_{u}$, and $\sigma_{g}$,
and in this regard they differ from the usual SSH model where $A$,
$B$, and $ C$ are constants. In order to vary these coefficients, it
is then necessary to vary the gate tension applied to the metallic
rods. Bloch's theorem allows us to write $\psi_{n}=\psi_{0}e^{iqnl}$
and $\xi_{n}=\xi_{0}e^{-iqnl}$. In matrix notation, Eq.~(\ref{eq:plasmonic_SSH equations}) then
becomes:

\begin{equation}
\left(\begin{array}{cc}
0 & A_{g}+A_{u}e^{-iql}\\
A_{g}+A_{u}e^{iql} & 0
\end{array}\right)\left(\begin{array}{c}
\psi_{0}\\
\xi_{0}
\end{array}\right)=C\left(\begin{array}{c}
\psi_{0}\\
\xi_{0}
\end{array}\right)
\end{equation}
The matrix with the coefficients $A_{g}$ and $A_{u}$ can be written as a linear
combination of the Pauli matrices $\sigma_{x}$ and $\sigma_{y}$,
as usual in these type of problems~\cite{2016short}. That is, we
can write
\begin{equation}
\begin{array}{c}
\left[A_{g}(\omega,d)+A_{u}(\omega){\rm cos}(ql)\right]\sigma_{x}+A_{u}(\omega){\rm sin}(ql)\sigma_{y}\\
\equiv d_{x}(\omega,d,q)\sigma_{x}+d_{y}(\omega,q)\sigma_{y}.
\end{array}\label{eq:coefficients_sigma_matrices}
\end{equation}
The coefficient $C$, playing a similar role of the eigenvalue on the SSH
model, has the form
\begin{equation}
C(\omega,q,d)=\pm\sqrt{A_{g}^{2}(\omega,d)+A_{u}^{2}(\omega)+2A_{g}(\omega,d)A_{u}(\omega){\rm cos}(ql)},\label{eq:C_equation}
\end{equation}
where the explicit dependencies of the coefficients on $\omega$,
$q$, and $d$ are kept as arguments of $A_{g}$, $A_{u}$, and $C$. The
two possible solutions of $C(\omega,q,d)$ correspond to two separate
bands when plotted against $ql$. This allows us to make a parallel
with the original SSH~\cite{2016short}, in which $C(\omega,q,d)$
plays the role of the SSH model's energy and Eq.~(\ref{eq:coefficients_sigma_matrices})
plays the role of the Hamiltonian. We make use of this parallelism
to study the topological features of the plasmonic SSH. 

\subsection{Topological characterization in momentum space}

\begin{figure}[t!]
	\begin{centering}
		\includegraphics[width=8.5cm]{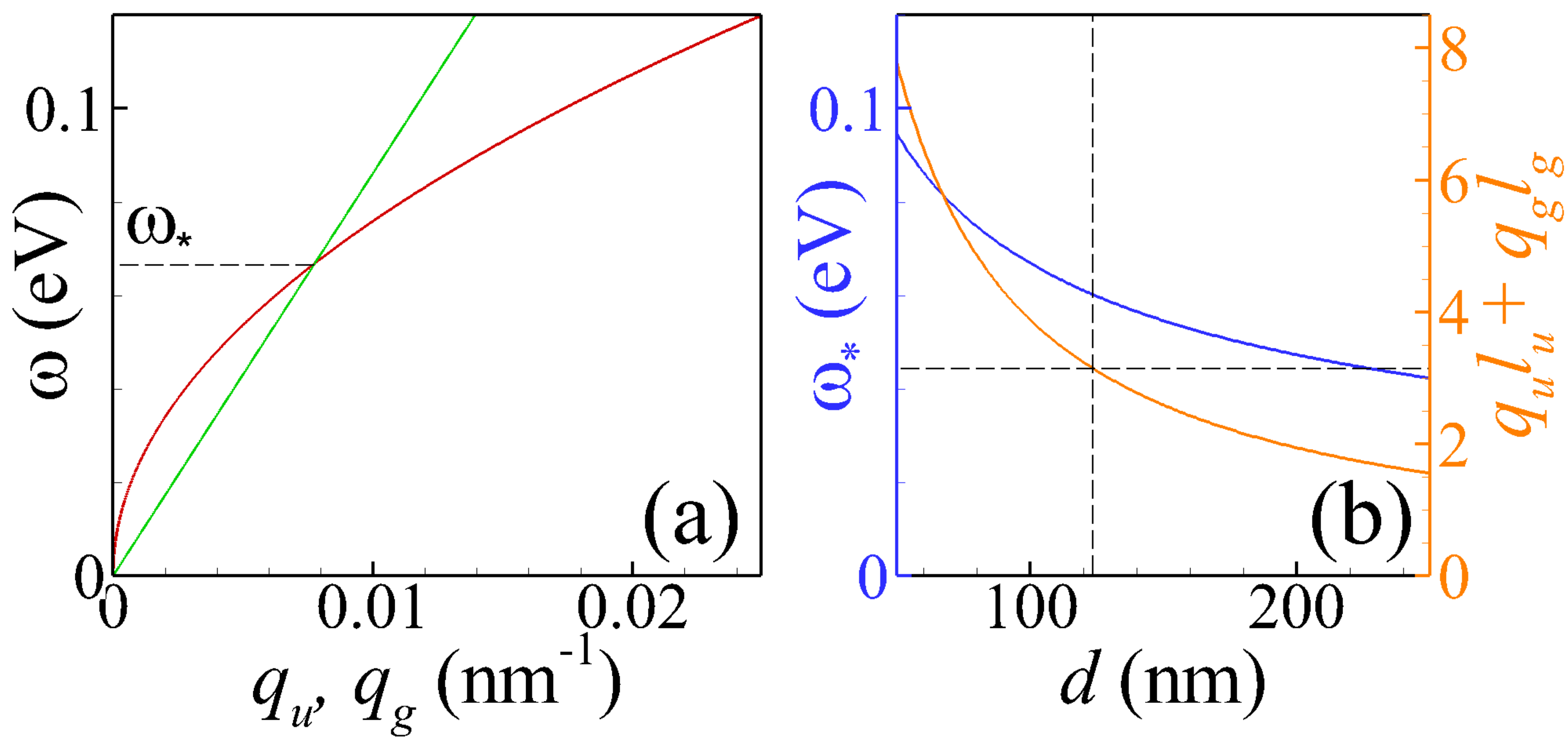}
		\par\end{centering}
	\caption{(a) Dispersion relations of gated [Eq.~\eqref{eq:q_gated}, green line] or ungated [Eq.~\eqref{eq:q_ungated}, red line] plasmons for $d=100$\,nm. Dashed horizontal line depicts frequency $\omega_*$, for which $q_u=q_g$; (b) Dependence of frequency $\omega_*$ (left vertical axis) and phase over period $q_ul_u+g_gl_g$ (right axis) upon distance $d$. Dashed horizontal line correspond to phase $q_ul_u+g_gl_g=\pi$, while vertical dashed line depicts critical thickness $d_c$. In all panels other parameters are: $E_F=0.45$\,eV, $\epsilon_1=1$, $\epsilon_2=3.5$, $l_{u}=260$\,nm, and $d=100$\,nm  .}
\label{fig:zero-gap}	
\end{figure}

\begin{figure*}
\begin{centering}
\includegraphics[scale=0.42]{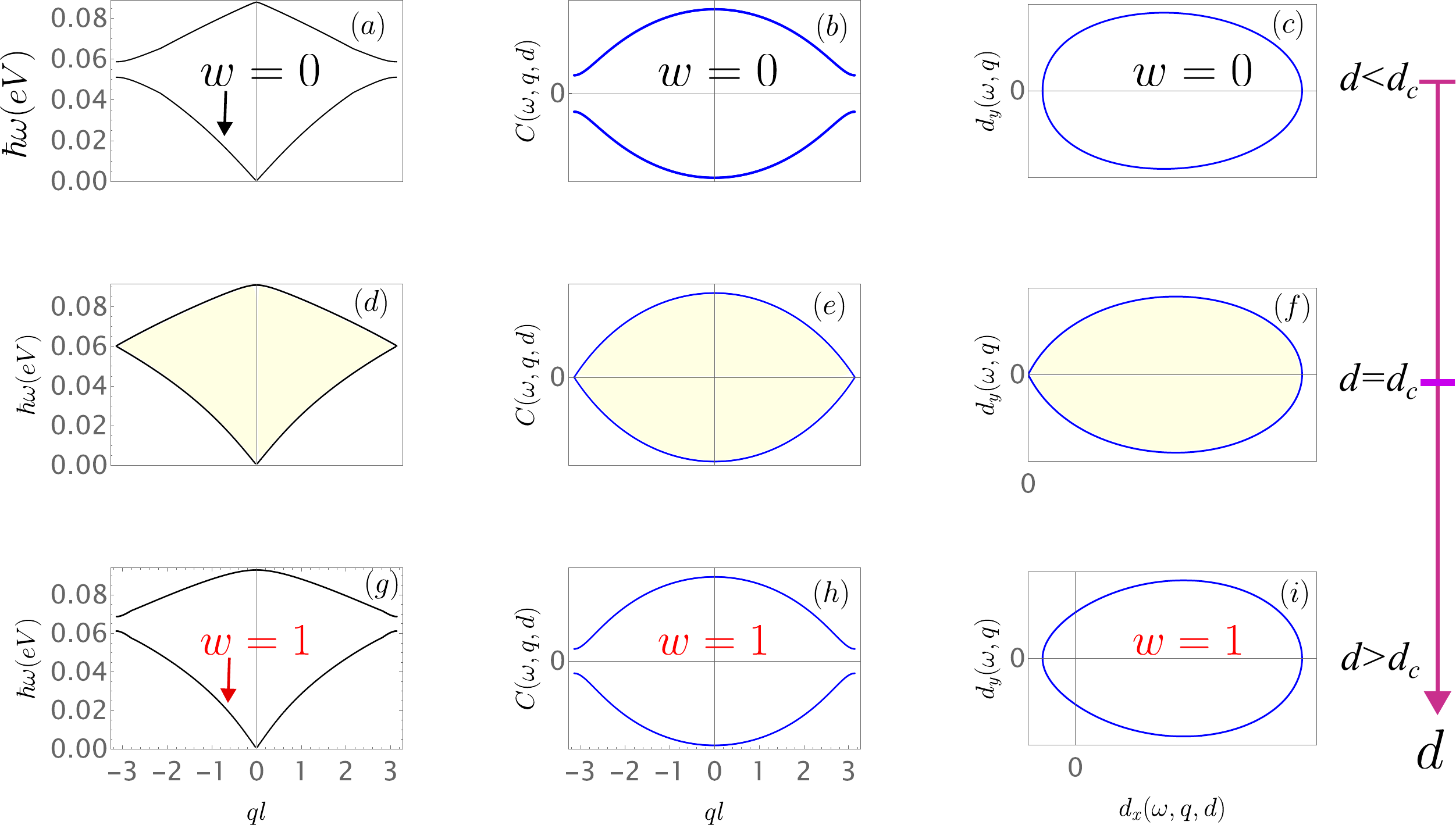}
\par\end{centering}
\caption{\label{fig:topological_characteristics_band_1}Topological characterization
of the first energy band of the plasmonic SSH crystal. The parameters
for this representations were $l=500$\,nm, $l_{g}=240$\,nm, $l_{u}=260$\,nm, $E_{F}=0.45$\,eV, $\epsilon_{1}=1$ and $\epsilon_{2}=3.5$. For
this choice of parameters, $d_{c}\approx123.79$\,nm, for which
the gap closes at $\hbar\omega\approx0.060$\,eV. The distance
to the metal was chosen to be $d=80$\,nm for the first row and $d=200$\,nm for the third row. ${\bf a)},{\bf d)},{\bf g)}$ Energy bands
as a function of $ql$; ${\bf b)},{\bf e)},{\bf h)}$ Coefficient
$C(\omega,q,d)$ as a function of $ql$; ${\bf c)},{\bf f)},{\bf i)}$
Parametric plot in the plane $d_{x}-d_{y}$ as $ql$ varies from $-\pi$
to $\pi$. We emphasize that for each $ql$, the corresponding value
of $\hbar\omega$ in the first energy band has to be computed into
$d_{x}(\omega,d,q),d_{y}(\omega,q)$. For $d<d_{c}$\,nm, the topology
of the first band is trivial. For $d=d_{c}$, we find ourselves at
the boundary between topological phases. For $d>d_{c}$, the topology
of the first band is non-trivial.}
\end{figure*}

In this section, we investigate the topology of our plasmonic SSH-line 
model, focusing our attention on the behavior of the first and second plasmonic
bands. We note that, when there is a gap in the energy spectrum (take for example, the first energy gap in Fig.~\ref{fig:KP_dispersion}), there is also a gap in the
plot $C(\omega,q,d)$ versus $ql$, i.e positive and negative solutions in Eq.~\eqref{eq:C_equation} are always different. As we vary the distance from the
metal $d$ we note that, as the first band gap closes, the gap in
$C(\omega,q,d)$ also closes at $ql=\pm\pi$, i.e positive and negative solutions in Eq.~\eqref{eq:C_equation} are equal for $ql=\pm\pi$. This behaviour is shown in the first two columns of Fig.~\ref{fig:topological_characteristics_band_1}. This is exact parallel of what
happens in the electronic SSH model, but with the energy eigenvalues $E(k)$ instead
of $C(\omega,q,d)$~\cite{2016short}. We then expect that, if $d_{c}$
is the distance a graphene from the metal that closes the energy gap and the
gap in $C(\omega,q,d)$, values of $d$ greater and smaller than $d_{c}$
will correspond to different topological phases. Thus, it would be valuable to find an analytical expression for $d_{c}$, as we do next for
the first energy band gap. Since both bands of $C(\omega,q,d)$ touch
at $ql=\pm\pi$ when the first energy gap closes, we can rewrite Eq.~(\ref{eq:C_equation}) as
\begin{equation}
C(\omega,\pm\frac{\pi}{l},d)=\left|A_{g}(\omega,d)-A_{u}(\omega)\right|=0
\end{equation}
Then it is clear that, for this condition to hold, $A_{g}=A_{u}$. By taking
the explicit form of the coefficients as in Eq.~(\ref{eq:definition_coefficients_ABC})
into consideration, the conditions $A_{g}=A_{u}$ and $C=0$ can be rewritten
as:
\begin{subequations}    
\begin{equation}
C=A_{g}{\rm cos}(q_{g}l_{g})+A_{g}{\rm cos}(q_{u}l_{u})=0\label{eq:C=00003D0}
\end{equation}
\begin{equation}
A_{g}=A_{u}=\frac{\sigma_{g}q_{g}}{{\rm sin}(q_{g}l_{g})}=\frac{\sigma_{u}q_{u}}{{\rm sin}(q_{u}l_{u})}\label{eq:A=00003DB}
\end{equation}
\end{subequations}
From Eq.~(\ref{eq:C=00003D0}), we see that the cosines are equal
but with opposite signs
\begin{equation}
{\rm cos}(q_{g}l_{g})=-{\rm cos}(q_{u}l_{u}).
\end{equation}
The sine terms ${\rm sin}(q_{g}l_{g})$ and ${\rm sin}(q_{u}l_{u})$ are then
either identical or have the same modulus but opposite sign as well.
From Eq.~(\ref{eq:A=00003DB}), we see that the sines must have the
same sign, since $\sigma_{g}q_{g}$ and $\sigma_{u}q_{u}$ are strictly
positive quantities and, to fulfill condition~\eqref{eq:A=00003DB}, they should satisfy
\begin{equation}
\sigma_{g}q_{g}=\sigma_{u}q_{u}.
\label{eq:qug}
\end{equation} 
Angles that have the same sine and opposite cosine
sum to an odd multiple of $\pi$:
\begin{equation}
q_{g}l_{g}+q_{u}l_{u}=m\pi,\quad m=1,3,5\ldots\label{eq:condition_odd_bands}
\end{equation}
Now, from Eqs.~\eqref{eq:qug} and \eqref{eq:condition_odd_bands} we can derive an analytical
expression for the distance $d_{c}$ from the metal that makes the two bands
of $C$ touch. This is straightforward
if the explicit expressions for $q_{g}$ and $q_{u}$ are considered, given
by Eqs.~(\ref{eq:q_gated}) and (\ref{eq:q_ungated}). Thus, for equal conductivities $\sigma_g=\sigma_u$, condition~\eqref{eq:qug} transforms into $q_u=q_g$, which is fulfilled for a certain frequency $\omega_*$ (such that $\hbar\omega_{*}$ is the energy at the point where the two bands touch) [see Fig.~\ref{fig:zero-gap}(a)]. By combining Eqs.~\eqref{eq:q_gated} and \eqref{eq:q_ungated}, an expression for frequency $\omega_*$ can be determined
\begin{equation}
\omega_*\left(d\right)=\sqrt{\frac{4E_{F}c\alpha \epsilon_{2}}{\hbar d(\epsilon_{1}+\epsilon_{2})^2}}.
\label{eq:omega_star}
\end{equation}
This frequency $\omega_*$ varies with thickness $d$ as $\sim 1/\sqrt{d}$, as it is shown in Fig.~\ref{fig:zero-gap}(b) by the blue line (left axis). At the same time, condition~\eqref{eq:condition_odd_bands} is fulfilled for $m=1$ and for critical thickness $d_c$ as expected [crossing of orange line in Fig.~\ref{fig:zero-gap}(b) with horizontal dashed line]. Substituting Eqs.~\eqref{eq:q_gated} and \eqref{eq:q_ungated} for frequency $\omega_*$ into condition~\eqref{eq:condition_odd_bands}, we obtain an equation for the critical thickness in the form 
\begin{equation}
\omega_*\left(d_c\right)\sqrt{\frac{\hbar\epsilon_{2}}{4E_{F}c\alpha d_{c}}}l_{g}+\frac{\hbar(\epsilon_{1}+\epsilon_{2})}{4E_{F}c\alpha}\left[\omega_*\left(d_c\right)\right]^2l_{u}=m\pi,
\label{eq:wstar}
\end{equation}
from which we obtain
\begin{equation}
d_{c}^{(m)}=\frac{\epsilon_{2}l}{m\pi(\epsilon_{1}+\epsilon_{2})}.\label{eq:d_close}
\end{equation}

\noindent This expression for $d_{c}$ is valid for all odd band gaps. By substituting the critical distance given by Eq. (\ref{eq:d_close}) into Eq. (\ref{eq:omega_star}), we find the energies for which the odd band gaps close as only a function of $m$ and the parameters of the system. We observe that if
$m=1$, we have the value of $\hbar\omega_*$ at which the first
band gap closes and, if $m=3$, we have the value of $\hbar\omega_*$ at which the third
band gap closes, and so on. At the
end, we see that $m$ plays the role of the band gap index. We would like to point out that
this formalism also applies to those band gaps that close at $ql=0$,
like the second and fourth ones. The only difference is that, for
this case, the cosines are equal in sign and modulus, while the sines
have equal modulus but opposite sign. We would arrive at the conclusion
that $q_{g}l_{g},q_{u}l_{u}$ sum to an even multiple of $\pi$, obeying
$q_{g}l_{g}+q_{u}l_{u}=p\pi$ with $p=2,4,6\ldots$. Thus, Eq.~(\ref{eq:d_close}) applies to all band gaps if we redefine $m=1,2,3,4\ldots$.

With the value of $d_{c}$ at hand, the remaining task is to identify
which topological phase corresponds to $d>d_{c}$ and which corresponds
to $d<d_{c}$ for the first and seconds plasmonic energy bands. This information can be obtained
with the functions $d_{x}$ and $d_{y}$ as defined in Eq.~(\ref{eq:coefficients_sigma_matrices}).
From the electronic SSH model~\cite{2016short} we know that, as $ql$
varies from $-\pi$ to $\pi$, these functions describe a curve in
the $d_{x}-d_{y}$ plane. This curve contains all information we need
to determine the topological phase the system is in, and it can be used to define the topological invariant associated with the system: the winding number $w$. If the curve
winds around the origin once, the system is said to be in the topological
phase and possesses winding number $w=1$; if the curve does not wind around the origin, the system is
said to be in the trivial phase and possesses winding number $w=0$; if the curve passes exactly through
the origin, the system lies between topological phases, the energy
band gap of the SSH model is closed and $w$ is not well defined. We expect that, after choosing
a value of $d$ and plotting $d_{x}(\omega,d,q)$ and $d_{y}(\omega,q)$
for the corresponding energies of the first and second energy bands, as $ql$
varies from $-\pi$ to $\pi$, we can find the topological invariant $w$ characterizing each bands. 

\begin{figure*}
\begin{centering}
\includegraphics[scale=0.42]{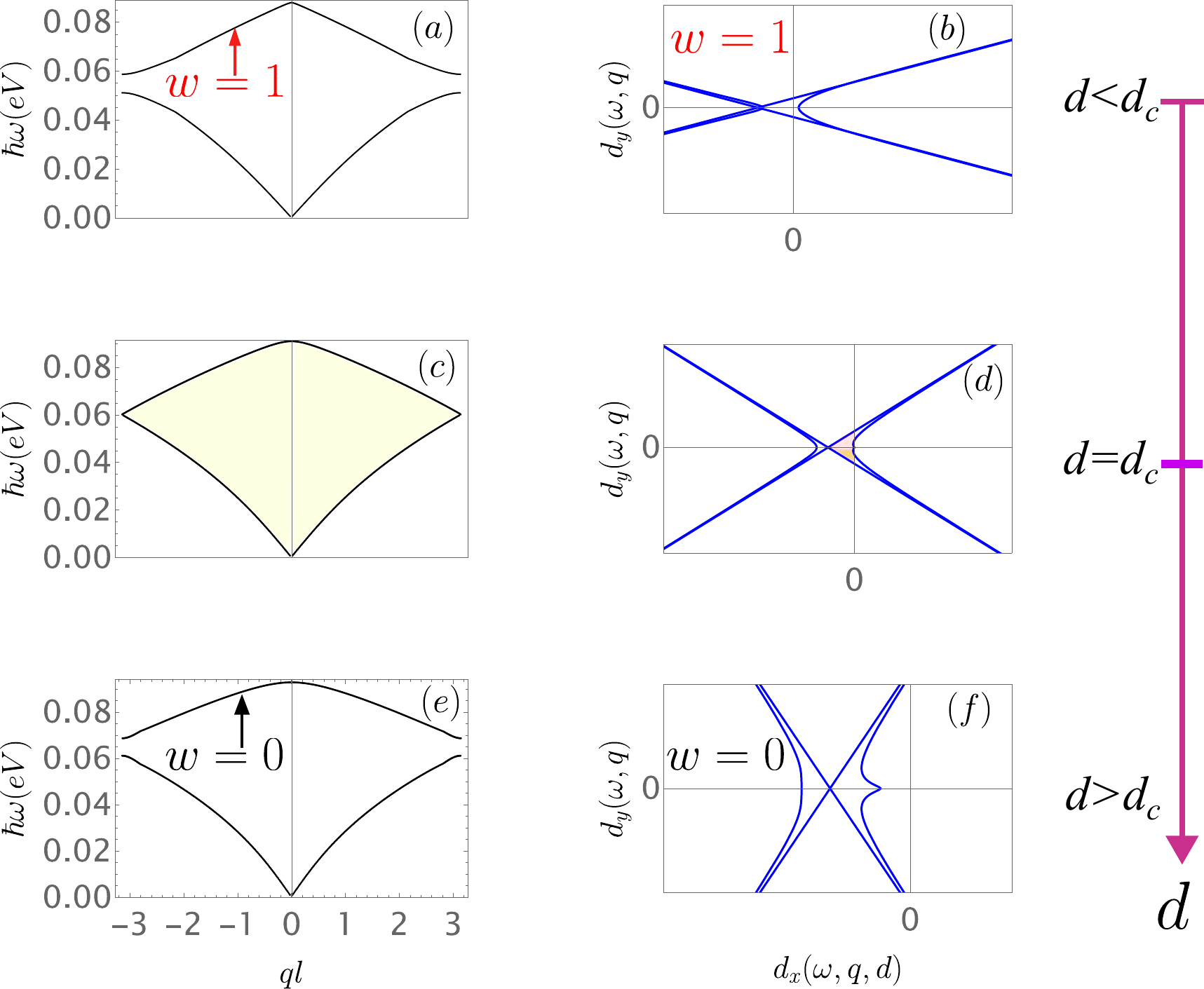}
\par\end{centering}
\caption{\label{fig:topological_characteristics_band_2}Topological characterization
of the second energy band of the plasmonic SSH crystal. The parameters
for this representations were $l=500$\,nm, $l_{g}=240$\,nm, $l_{u}=260$\,nm, $E_{F}=0.45$\,eV, $\epsilon_{1}=1$ and $\epsilon_{2}=3.5$. For
this choice of parameters, $d_{c}\approx123.79$\,nm, for which
the gap closes at $\hbar\omega\approx0.060$\,eV. The distance
to the metal was chosen to be $d=80$\,nm for the first row and $d=200$\,nm for the third row. ${\bf a)},{\bf c)},{\bf e)}$ Energy bands
as a function of $ql$; ${\bf b)},{\bf d)},{\bf f)}$
Parametric plot in the plane $d_{x}-d_{y}$ as $ql$ varies from $-\pi$
to $\pi$. For $d<d_{c}$\,nm, the topology
of the second band is non-trivial. For $d=d_{c}$, we find ourselves at
the boundary between topological phases. For $d>d_{c}$, the topology
of the second band is trivial. The situation is opposite in comparison to the topological characterization of the first plasmonic band in Fig.~\ref{fig:topological_characteristics_band_1}.}
\end{figure*}

The results of this discussion are shown in Fig.~\ref{fig:topological_characteristics_band_1} and in Fig.~\ref{fig:topological_characteristics_band_2}.
In Fig.~\ref{fig:topological_characteristics_band_1}, the first row corresponds to $d<d_{c}$, the second row to $d=d_{c}$,
and the third row to $d>d_{c}$. In the first column, the first two
energy bands are depicted. In the second column, the coefficient $C(\omega,q,d)$
is plotted against $ql$. In the third column, we plot the parametric
curve in the $d_{x}-d_{y}$ plane, as $ql$ varies from $-\pi$ to $\pi$. Initially, looking at only the
energy bands and the coefficient $C(\omega,q,d)$, it is not possible
to distinguish between the two gapped phases in the first and third row. However, the $d_{x}-d_{y}$ curve in the
third column for $d<d_{c}$ does not wind around the origin, while
the curve for $d>d_{c}$ does. This indicates that the first band
has trivial topology for $d<d_{c}$ (winding number $w_1=0$) and non-trivial topology for $d>d_{c}$ (winding number $w_1=1$).
In the second row, we use $d=d_{c}$ to show that indeed the energy
gap and the $C(\omega,q,d)$ gap close simultaneously. Additionally,
the corresponding $d_{x}-d_{y}$ curve passes exactly at the origin, confirming that the topological phase transition occurs for $d=d_{c}$. 
In Fig.~\ref{fig:topological_characteristics_band_2} we again depict the plasmonic energy bands in the first column. In the second column, we show the parametric curve in the $d_{x}-d_{y}$ plane. Interestingly, the situation for the second plasmonic band is reversed in comparison to the first band. For $d<d_{c}$, the curve in the $d_{x}-d_{y}$ plane winds around the origin once (winding number $w_2=1$) , while it does not wind around the origin for $d>d_{c}$ (winding number $w_2=0$). This indicates that when the first plasmonic band has trivial topology, the second plasmonic band has non-trivial topology and vice-versa. Then, for a certain $d$, the difference between topological invariants between the first and second bands is always $w_2-w_1=\pm1$, and this guarantees the existence of one edge mode in the first energy gap of the finite plasmonic crystal (see discussion below).

\subsection{End states}

\begin{figure}
\begin{centering}
\includegraphics[scale=0.45]{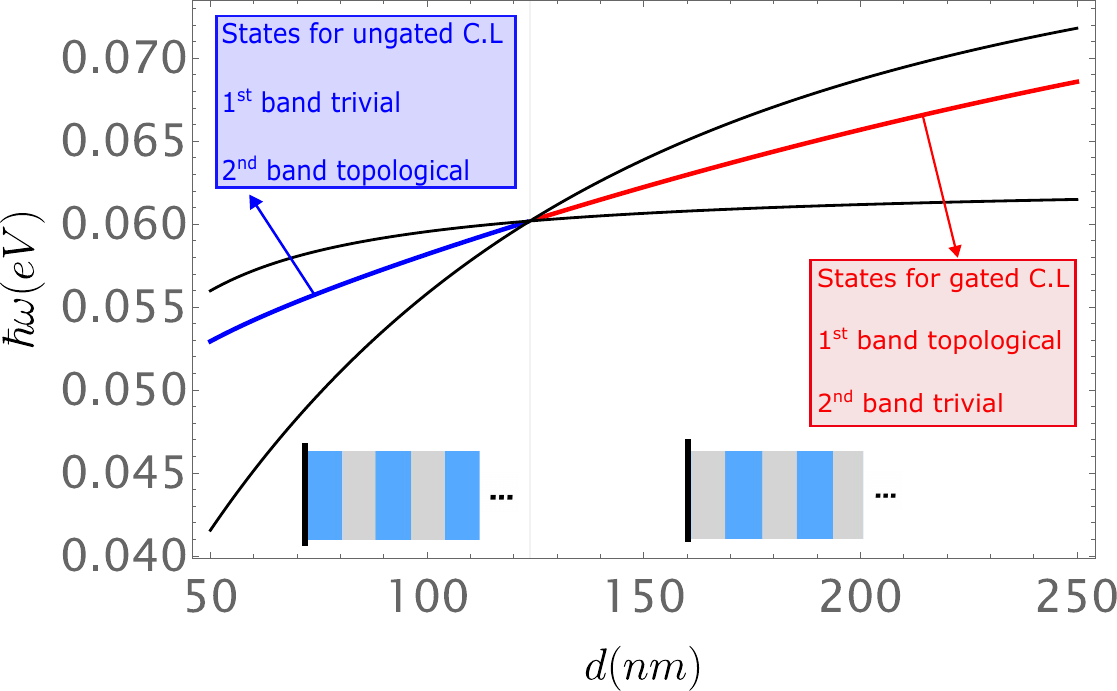}
\par\end{centering}
\caption{\label{fig:gap_states_band1}End states in the first energy gap of the infinite crystal as function of the graphene to metal distance $d$. The black curves are band edges, located at $ql=\pm\pi$. When the gap closes, the band edges touch and the end states vanish. We note that there are edge state both to the left (blue line) and to the right (red line) of the critical value of the graphene to metal distance $d_c$. Studying the topology of the two bulk first bands, we find that to the left of $d_c$ the first band in trivial and the second is topological, leading to a difference $\Delta w_1=+1$, whereas to the right of $d_c$ the opposite happens and $\Delta w_1=-1$. In both cases we have topological end states present in the semi-infinite crystal.}
\end{figure}

After investigating the topology of the first two bands in the infinite crystal, 
the  next step forward is to look for the semi-infinite case in search of end states. This can be done by chopping the infinite plasmonic crystal at a given location, giving rise to an edge. The semi-infinite crystal is defined to the right of the edge.  The crystal can then  begin either at an ungated or at a gated region, which we call capping layer (C.L) from now on. From what is known on topological systems in one dimension, like the SSH electronic model, localized states are expected to appear at the terminations of the system for parameters that drive the bulk to the topological phase. This phenomenon is known as bulk-edge correspondence \citep{Hatsugai1993,Mong2011}, through which we can relate the number of localized end states per edge (which appear inside the gap of the infinite system spectrum) with the value of the topological invariant in the bulk \citep{2016short}. We note that this is exactly what happens here, as shown in Fig.~\ref{fig:gap_states_band1}, where we plot the band edges (black) as a function of $d$. If the system begins with an ungated capping layer, end states only appear inside the gap for values of $d$ that make the first band trivial and the second band topological in the infinite crystal, that is, $d<d_{c}$. On the other hand, if the system begins with a gated capping layer, end states only appear for values of $d$ that make the first band topological and the second band trivial in the infinite case, that is, $d>d_{c}$. As mentioned before, the difference between topological invariants of the first and second bands is $w_2-w_1=\pm1$, which guarantees that the gap between these bands always contains an end state by the bulk-edge correspondence. This state has complex Bloch momentum $q$, with $\rm{Re}[q]=\pm\pi$ and it must have a positive imaginary part, which guarantees that it is decaying inside the semi-infinite crystal. If we consider larger energy scales and bands with higher indexes ($m=3,4,5\dots$) there may appear other end states, which would live inside higher band gaps, but since we are only interested in the first two bands, the only end states that interest us are the ones appearing inside the first and second band gaps.

There is one additional study we must perform, namely the existence of end states as function the ratio $l_u/l$, with $l_u$ and $l$ being the length of ungated regions and the length of the unit cell, respectively.

\begin{figure}

\begin{minipage}[t]{0.45\textwidth}%
\begin{center}
\includegraphics[scale=0.85]{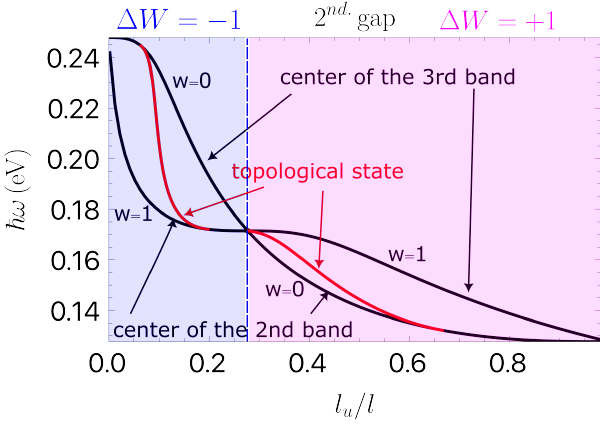}
\par\end{center}%
\end{minipage}\hspace{0.2\textwidth}%
\begin{minipage}[t]{0.45\textwidth}%
\begin{center}
\includegraphics[scale=0.85]{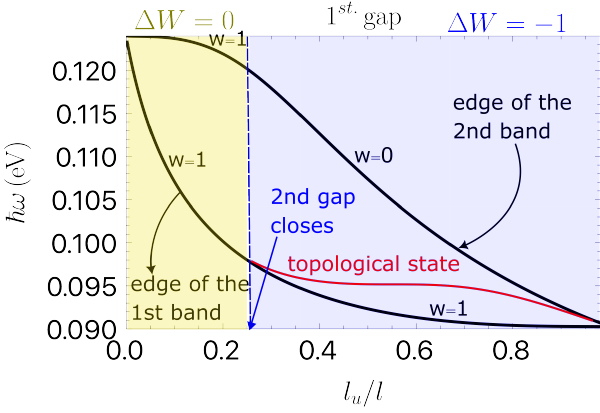}
\par\end{center}%
\end{minipage}

\caption{\label{fig:gap_as_fun_lu}
Topological and trivial gaps as function of the length $l_u$ of the ungated region. We see formation of end states in the gaps when the winding number difference, $\Delta w_m$, is different from zero (both three cases $\Delta w=0,\pm 1$ are observed). We note the peculiar situation of the change of topological nature of the second band near the band edge (bottom panel) when the gap between this band the third band (center of the zone) closes. 
The
parameters used to produce these figures were $l=500$\,nm, $E_{F}=0.45$\,eV, $\epsilon_{1}=1$, $\epsilon_{2}=1$, and $d=150$ nm.  Similar results were found for topological Tamm states \citep{Henriques_2020}. (For a complete understanding of the terminology used in this figure, the reader is referred to Fig.~\ref{fig:KP_dispersion}.)}
\end{figure}

For a two band system, as the original electronic SSH model, gap states can be easily understood, since they can only appear in one gap and there are only two topological invariant $w_m$ ($m=\,1,2$). In a multi-band one-dimensional system the situation is more complex, as each band, except for the first one, is sandwiched between two gaps. Then we may face the situation where the winding number of the bands change due to gaps above and bellow it opening and closing. This situation is considerably more complex, as depicted in Fig.~\ref{fig:gap_as_fun_lu}, where the winding number of the second band is changing from $w=1$ to $w=0$, even though the first gap stays open (at the band edge). Instead, the winding number of the second band changes because the gap between the second and the third bands (at $k=0$) closes.  This shows the intricacy of the study of topology in a multi-band system with several gaps opening and closing independently. Fig.~\ref{fig:gap_as_fun_lu} summarizes well what is  discussed in this section.

Once the existence of end states (and its topological nature) is fixed, the question about their observation arises. Below, we propose that a reflection experiment, where a plasmon wave impinges on a finite plasmonic crystal coupled to a thin metallic film, will show a clear  signature of the 
presence of topological decaying end states
in the reflection coefficient.

\begin{figure}
\begin{centering}
\includegraphics[scale=0.6]{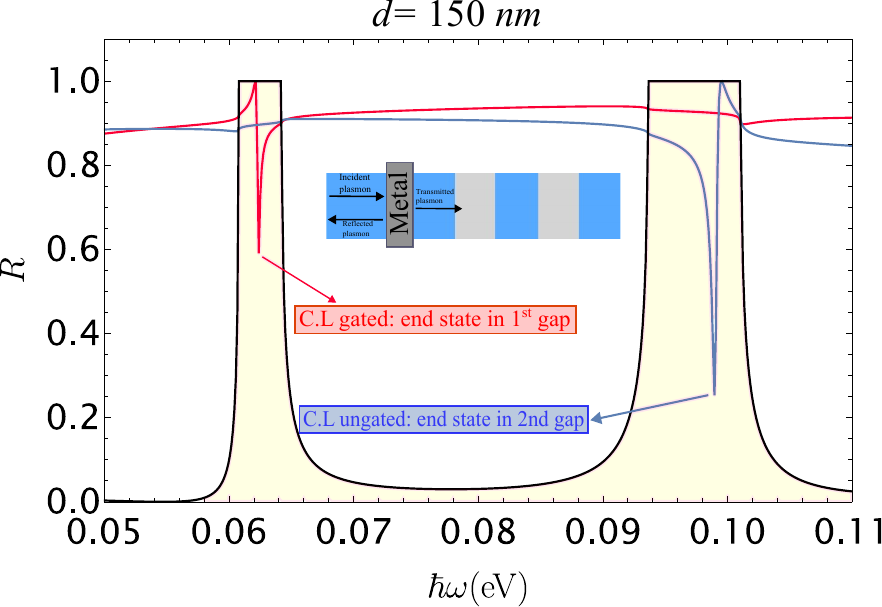}
\par\end{centering}
\caption{\label{fig:reflectance} Reflectance of an incoming plasmon backscattered by a semi-infinite plasmonic crystal with a metallic layer. An aluminium film was considered, with $2$ nm thickness and $1$ nm length. Properties of the aluminium film were taken from \citep{Rakic1998}. The semi-infinite crystal parameters were $l=500$\,nm, $l_{g}=240$\,nm, $l_{u}=260$\,nm, $E_{F}=0.45$\,eV, $d=150$ nm, $\epsilon_{1}=1$ and $\epsilon_{2}=3.5$. The black curve bounding the yellow region is the reflectance spectrum of a semi-infinite crystal. The red and blue curves correspond to the reflection on an incoming plasmon by the metal plus the semi-infinite plasmonic crystal. The existence of end states in the semi-infinite crystal is signalled by a dip in the reflectance spectrum. The capping layer of the semi-infinite crystal that borders the metallic layer to the left determines whether end states will exist for a certain $d$. States exist in the first gap if the capping layer is a gated region or in the second gap if the capping layer is an ungated region. Changing $d$ may change the topological invariant of the bands and thus states may or may not continue to exist in either gap. The inset shows an schematic of the semi-infinite plasmonic crystal with a metallic layer before it, in which the capping layer to the right of the metal is ungated.}
\end{figure}

To avoid the well known oscillations in the reflection coefficient typical of finite crystals, our results were obtained for a semi-infinite crystal, such that there is no reflected wave coming back from infinity. The bottom line is that the reflection coefficient shows a dip inside the bulk energy gaps whenever a topological state exists in the gap and is flat in the opposite case. In Fig.~\ref{fig:reflectance} we show the reflectance spectra of the infinite crystal (black) alongside with two curves (blue and red), corresponding to different choices of capping layers. The plasmons impinge from the left into the metallic film, that replaces graphene in that zone, as shown in the inset of Fig.~\ref{fig:reflectance}. To the right of the metal, the semi-infinite plasmonic crystal will support the localized end states and also Bloch modes. The component that is reflected back at the metallic interface will lack the end state frequency. Depending on the kind of capping layer, end states may or may not exist for a certain choice of $d$, as is shown in Fig.~\ref{fig:gap_states_band1}. The choice of $d=150$ nm makes the first band topological for a gated capping layer and the second band topological for an ungated capping layer. Thus, we expect that for a gated capping layer, no dip will be seen in the second band, while for an ungated capping layer, no dip will be seen in the first gap. This is exactly what Fig.~\ref{fig:reflectance} shows, suggesting that plasmonic end states can indeed be detected experimentally.

Finally, we outline below the calculation of the reflectance amplitude of a semi-infinite crystal ($x>0$) connected to a semi-infinite ($x<0$) encapsulated graphene sheet, that supports plasmons coming from $-\infty$.
We write the passage of the plasmon from the $x<0$ to the $x>0$ regions as 
 
 \begin{equation}
 \left[\begin{array}{c}
	\phi_{\alpha+}^{1}\\
	\\
	\phi_{\alpha-}^{1}
\end{array}\right]={\mathbf M}_{\alpha}\left[\begin{array}{c}
	1\\
	\\
	r
\end{array}\right]\equiv \left[\begin{array}{cc}
	m_{11,\alpha} & m_{12,\alpha}\\
	\\
	m_{21,\alpha} & m_{22,\alpha}
\end{array}\right]\left[\begin{array}{c}
	1\\
	\\
	r
\end{array}\right]. 	
 \end{equation}	
where $\alpha=q,u$, depending on whether the semi-infinite crystal begins with a gated or ungated capping layer. Thus $\phi_{\alpha+}^{1}	=m_{11,\alpha}+m_{12,\alpha}r$ and 
$\phi_{\alpha-}^{1}	=m_{21,\alpha}+m_{22,\alpha}r$. 
Dividing the last two equations by each other we find
\begin{equation}
	\frac{\phi_{\alpha-}^{1}}{\phi_{\alpha+}^{1}}=\frac{m_{21,\alpha}+m_{22,\alpha}r}{m_{11,\alpha}+m_{12,u}r}
\end{equation}
Now, from Bloch's theorem: 
 \begin{equation}
 e^{iql}\left[\begin{array}{c}
	\phi_{\alpha+}^{1}\\
	\\
	\phi_{\alpha-}^{1}
\end{array}\right]={\mathbf M}\left[\begin{array}{c}
	\phi_{\alpha+}^{1}\\
	\\
	\phi_{\alpha-}^{1}
\end{array}\right]\equiv \left[\begin{array}{cc}
	m_{11} & m_{12}\\
	\\
	m_{21} & m_{22}
\end{array}\right]\left[\begin{array}{c}
	\phi_{\alpha+}^{1}\\
	\\
	\phi_{\alpha-}^{1}
\end{array}\right]. 	
 \end{equation}	
where $\mathbf M$ is the transfer matrix of a unit cell of the semi-infinite crystal. We can obtain another relation for the ratio $\frac{\phi_{\alpha-}^{1}}{\phi_{\alpha+}^{1}}$ 
\begin{equation}
	\frac{\phi_{\alpha-}^{1}}{\phi_{\alpha+}^{1}}=\frac{e^{iql}-m_{11}}{m_{12}}=\frac{m_{21,\alpha}+m_{22,\alpha}r}{m_{11,\alpha}+m_{12,u}r}\equiv\rho(q).
\end{equation} 
The term $e^{iql}$ requires the determination of Bloch wave vector $ql$, which can be easily obtained from the plasmonic version of the Kronig-Penney equation in the form $ql=\pm\arccos[{\rm Tr}(\mathbf{M})/2]$. We have therefore two solutions for $ql$. The correct solution is the one satisfying the condition $\vert\rho(q)\vert\le1$. Solving the previous equation for $r$ we find
\begin{equation}
	r=\frac{\rho(q)m_{11,\alpha}-m_{21,\alpha}}{m_{22,\alpha}-m_{12,\alpha}\rho(q)},
\end{equation}
The black curve in Fig.~\ref{fig:reflectance} is $R=\left|r\right|^{2}$.
The calculation of the red and blue reflectance curves in Fig.~\ref{fig:reflectance}
are obtained by generalizing this approach to the existence of a metallic layer. Now, one must relate the wave amplitudes at the beginning (left) and at the end (right) of the metallic layer using a transfer matrix ${\mathbf M}_{\rm{metal}}$, which contains information about the metallic film. The amplitudes at the end of the metallic layer can then be related with the bulk of the semi-infinite crystal through Bloch's theorem and the unit cell transfer matrix $\mathbf M$. The red curve was obtained considering a gated capping layer, such that $\mathbf M=\mathbf{M}_u\mathbf{M}_g$, while the blue curve was obtained using an ungated capping layer $\mathbf M=\mathbf{M}_g\mathbf{M}_u$. The inset in Fig.~\ref{fig:reflectance} shows an schematic of the semi-infinite plasmonic crystal with a metallic layer before it, in which the capping layer to the right of the metal is ungated.

\subsection{Analytical approximation for Dirac bands}

Another interesting aspect of the plasmonic SSH model is the fact
that, when the energy gaps close, we obtain Dirac cones. These ocour both at the center and at the edge of the plasmonic crystal Brillouin zone. It is possible
to obtain an analytical approximations for the bands by inspecting
the relationship between coefficients $A_{g}$ and $A_{u}$. As stated in the previous section, for odd band gaps we have 
\begin{equation}
\begin{array}{c}
A_{g}=A_{u}=\frac{\sigma_{g}q_{g}}{{\rm sin}(q_{g}l_{g})}=\frac{\sigma_{u}q_{u}}{{\rm sin}(q_{u}l_{u})}\\
\\
{\rm sin}(q_{g}l_{g})={\rm sin}(q_{u}l_{u})
\end{array}
\end{equation}
meaning that $\sigma_{g}q_{g}=\sigma_{u}q_{u}$. This is also true
for even band gaps, since in these cases the conditions are
\begin{equation}
\begin{array}{c}
A_{g}=-A_{u}=\frac{\sigma_{g}q_{g}}{{\rm sin}(q_{g}l_{g})}=-\frac{\sigma_{u}q_{u}}{{\rm sin}(q_{u}l_{u})}\\
\\
{\rm sin}(q_{g}l_{g})=-{\rm sin}(q_{u}l_{u})
\end{array}
\end{equation}
Thus, the relationship $\sigma_{g}q_{g}=\sigma_{u}q_{u}$ is always
true, given that $q_{g},q_{u}$ are evaluated at the energy where
the bands touch and at $d=d_{c}$. In this scenario, we can alter
the form of the Kronig--Penney-like dispersion found for the graphene
plasmons in Eq.~(\ref{eq:K-P plasmon dispersion}) by substituting
$\sigma_{g}q_{g}=\sigma_{u}q_{u}$ 

\begin{equation}
\begin{array}{c}
\begin{array}{c}
{\rm cos}(ql)={\rm cos}(q_{u}l_{u}){\rm cos}(q_{g}l_{g})-{\rm sin}(q_{u}l_{u}){\rm sin}(q_{g}l_{g})\end{array}\\
={\rm cos}(q_{u}l_{u}+q_{g}l_{g})
\end{array}
\end{equation}
The general solution to this equation has the form
\begin{equation}
\pm ql+2n\pi=\frac{\hbar(\epsilon_{1}+\epsilon_{2})\omega^{2}}{4E_{F}c\alpha}l_{u}+\omega\sqrt{\frac{\hbar\epsilon_{2}}{4E_{F}c\alpha d_{c}}}l_{g},
\end{equation}
where $n\in\mathbb{Z}$ and the explicit expressions for $q_{u}$
and $q_{g}$ were used. Solving the quadratic equation for $\hbar\omega$
and preserving only the solutions corresponding to positive energies
we arrive at
\begin{widetext}
\begin{equation}
\hbar\omega=\frac{l_{g}\sqrt{E_{F}\hbar c\alpha\epsilon_{2}}}{l_{u}(\epsilon_{1}+\epsilon_{2})\sqrt{d_{c}}}\left(\sqrt{1+\frac{4d_{c}}{\epsilon_{2}}\frac{(\epsilon_{1}+\epsilon_{2})l_{u}(\pm ql+2n\pi)}{l_{g}^{2}}}-1\right).\label{eq:approx_energy}
\end{equation}
\end{widetext}
Again, we emphasize that $\frac{\sigma_{u}q_{u}}{2\sigma_{g}q_{g}}+\frac{\sigma_{g}q_{g}}{2\sigma_{u}q_{u}}$
= 1 only at the point where the bands touch and thus Eq.~(\ref{eq:approx_energy})
is only an approximation. At the vicinity of the first gap (which is closed in this case) we use an expression for the critical thickness~\eqref{eq:d_close} for $m=1$, as well as define Bloch wavevectors in the limits $0\le ql \le 2\pi$. In this situation, two branches of frequencies can be obtained from Eq.~\eqref{eq:approx_energy} as
\begin{eqnarray}
\hbar\omega=\hbar\omega_*\left(d_c^{(1)}\right)\frac{\sqrt{l_g^2+\frac{ql}{\pi}4ll_u}-l_g}{2l_u},\label{eq:approx_1_1}\\
\hbar\omega=\hbar\omega_*\left(d_c^{(1)}\right)\frac{\sqrt{l_g^2+\left(2-\frac{ql}{\pi}\right)4ll_u}-l_g}{2l_u}.\label{eq:approx_1_2}
\end{eqnarray}  
At the vicinity of the second gap, the expression for the critical thickness~\eqref{eq:d_close} should be used with $m=2$, and the wavevector is defined inside the limits $-\pi\le ql\le\pi$. In this case, Eq.~\eqref{eq:approx_energy} for the two branches of frequencies can be rewritten as 
\begin{eqnarray}
\hbar\omega=\hbar\omega_*\left(d_c^{(2)}\right)\frac{\sqrt{l_g^2+\left(1\pm\frac{ql}{2\pi}\right)4ll_u}-l_g}{2l_u}.\label{eq:approx_2}
\end{eqnarray}
Close to the Dirac points, the square root
in the expression above can be Taylor expanded such that the relationship
between $\omega$ and $ql$ is linear. For the first Dirac point, Eqs.~\eqref{eq:approx_1_1} and \eqref{eq:approx_1_2} in the vicinity of $ql=\pi$ will be represented as
\begin{equation}
\hbar\omega=\hbar\omega_*\left(d_c^{(1)}\right)\left[1\pm\left(\frac{ql}{\pi}-1\right)\frac{l}{l+l_u}\right].
\end{equation}
For the second Dirac point, Eq.~\eqref{eq:approx_2} in the vicinity of $ql=0$ can be expanded as 
\begin{equation}
\hbar\omega=\hbar\omega_*\left(d_c^{(2)}\right)\left(1\pm\frac{ql}{2\pi}\frac{l}{l+l_u}\right).
\end{equation}
This concludes the derivation of the linear spectrum when the gap closes both at the edge and at the center of the Brillouin zone.

\section{Conclusions}

In this work we investigated a graphene-based polaritonic crystal.
We showed that the plasmon energies obey a Kronig--Penney dispersion
and calculated their propagation on the crystal with the transfer-matrix formalism. Upon realization that the crystal possesses sub-lattice
symmetry, we managed to derive tight-binding equations for the electrostatic
potential $\phi(x)$ describing the plasmons. At the end, these equations
resemble the SSH model tight-binding, one difference being that the
hopping amplitudes are energy dependent. This representation is convenient
because it allows us to bring the traditional formalism used to describe
topological insulators in one dimension to plasmonics. By investigating
the energy dispersion and the tight-binding coefficients we demonstrate
that, by tuning the distance $d$ between graphene and the metal grating,
it is possible to associate trivial and non-trivial topology to the
energy bands of the polaritonic crystal. Additionally, we derived
an analytical expression for $d_{c}$, the value of $d$ that closes
the energy band gaps. This expression can be used alongside numerical
methods to easily identify the boundary between topological phases
and to simplify the calculations involving the energy-dependent hopping
amplitudes. Lastly, an analytical approximation for the energy
bands was obtained by slightly modifying the Kronig--Penney-like dispersion.
We show that this approximation describes the energy bands well if
the energy gaps are closed, specially around Dirac points.  

We believe
the findings of our work have the potential to facilitate future studies
on topological plasmonics and provide a reliable framework to investigate
similar polaritonic crystals analytically. 

Also, it would be interesting to formulate the problem in terms of Wannier functions, \cite{Pedersen1991,Gupta2022}  a line of research we plan to pursue in the near future.

\section*{Acknowledgements }

D. A. M. acknowledges the project PTDC/FIS-MAC/2045/2021 for a research
grant and the hospitality of  the Center for Polariton-driven Light--Matter
Interactions (POLIMA),  funded by the Danish National Research Foundation
(Project No. DNRF165). N. M. R. P. acknowledges support by the Portuguese
Foundation for Science and Technology (FCT) in the framework of the
Strategic Funding UIDB/04650/2020, COMPETE 2020, PORTUGAL 2020, FEDER,
and through project PTDC/FIS-MAC/2045/2021. N. M. R. P. also acknowledges the hospitality of the POLIMA Center and, together with N. A. M., 
the Independent Research Fund Denmark (grant no. 2032-00045B) and
the Danish National Research Foundation (Project No. DNRF165).

\appendix
\section{Approximations for gated and ungated momenta }

In the non-retarded limit the electromagnetic fields in graphene and surrounding media are defined by Poisson equation
\begin{equation}
\epsilon_0{\rm div}\left[\epsilon\left(z\right){\rm grad} \phi\left(x,z\right)\right]=-\rho_{2D}\delta\left(z\right),
\label{eq:Poisson}
\end{equation}
where $\phi\left(x,z\right)$ is the scalar potential, 
\begin{equation}
\epsilon\left(z\right)=\left\{ \begin{array}{c}
\epsilon_{1},\qquad z<0\\
\epsilon_{2}, \qquad z>0
\end{array}\right.
\end{equation} is the dielectric permeability, and $\rho_{2D}$ is the two-dimensional charge density in graphene. In the right-hand part of Eq.\,\eqref{eq:Poisson}, the delta-function describes the fact that charges in graphene are concentrated on a two-dimensional plane, arranged at $z=0$. Beyond the graphene plane, the Poisson equation can be solved separately in the spatial domains $z>0$ and $z<0$, where the dielectric permeabilities are homogeneous, and the solution will have the form
\begin{equation}
\phi\left(x,z,t\right)=\left\{ \begin{array}{c}
\phi\left(0,0,0\right) e^{iq_{u}x-i\omega t}e^{q_{u}z},\qquad z<0\\
\phi\left(0,0,0\right) e^{iq_{u}x-i\omega t}e^{-q_{u}z},\qquad z>0
\end{array}\right.
\label{eq:particular-solutions}
\end{equation}
Here the scalar potential is continuous at $z=0$, and chose of signs in the exponential terms, containing $\pm q_{u}z$ guarantees that the surface wave is evanescent in the $z$ direction, and decays with increasing distance from the graphene plane at $z=0$. Also solution \eqref{eq:particular-solutions} implies that surface plasmons propagate along the $x$-axis with the wavevector $q_{u}>0$ and is oscillating in time with frequency $\omega$.

If the Poisson equation \eqref{eq:Poisson} is integrated along $z$-coordinate in the limits $z\in\left[-0,+0\right]$, it transforms into
\begin{equation}
\epsilon_2\left.\frac{\partial \phi}{\partial z}\right|_{z=+0}-\epsilon_1\left.\frac{\partial \phi}{\partial z}\right|_{z=-0}=-\frac{\rho_{2D}}{\epsilon_0}. 
\label{eq:Poisson-integrated}
\end{equation}
Substituting Eq.\,\eqref{eq:particular-solutions} into Eq.\,\eqref{eq:Poisson-integrated}, we obtain
\begin{equation}
q_u\left(\epsilon_2+\epsilon_1\right)\phi\left(0,0,0\right)e^{iq_{u}x-i\omega t}=\frac{\rho_{2D}}{\epsilon_0}. 
\end{equation}
The correspondent two-dimensional charge density can be obtained from the continuity equation
\begin{equation}
\frac{\partial \rho_{2D}}{\partial t}\delta\left(z\right) + {\rm div}\,{\bf j}=0, 
\label{eq:continuity-eq}
\end{equation}
where three-dimensional current {\bf j} in our particular case possesses only $x$-component \begin{equation}
j_x=-\sigma\left(\omega\right)\delta\left(z\right)\frac{\partial \phi}{\partial x}.
\label{eq:jx}
\end{equation}
where $\sigma(\omega)=\frac{4iE_{F}}{(\hbar\omega+i\Gamma)}\alpha\epsilon_{0}c$
is the Drude 2D conductivity of graphene, the dominant conductivity term from THz
to mid-IR \cite{GraphenePlasmons}. Here $E_F$ and $\Gamma$ stand for the Fermi energy and inverse scattering time, respectively. In this paper we consider the case, where losses in graphene are absent ($\Gamma\equiv 0$). Substituting current density from Eq. \eqref{eq:jx} into the continuity equation \eqref{eq:continuity-eq} and integrating it in time, we obtain the 2D charge density
\begin{equation}
\rho_{2D}=\frac{q_u^2\sigma\left(\omega\right)}{i\omega}\phi\left(0,0,0\right) e^{iq_{u}x-i\omega t}.
\label{eq:rho-2D}
\end{equation}
This equation, being substituted into Eq.\,\eqref{eq:Poisson-integrated}, gives the dispersion relation for ungated plasmons
\begin{equation}
q_{u}=\frac{\hbar(\epsilon_{1}+\epsilon_{2})\omega^{2}}{4cE_{F}\alpha}.
\end{equation}
If metal is present at a distance $d$, the scalar potential turns put to be more complicated
\begin{eqnarray}
\phi\left(x,z,t\right)=\phi\left(0,0,0\right) e^{iq_gx-i\omega t}\times\nonumber\\
\left\{ \begin{array}{c}
e^{q_gz},\qquad z<0\\
\frac{{\rm sinh}(q_g[d-z])}{{\rm sinh}(q_gd)},\qquad 0<z<d\\
0,\qquad z>d
\end{array}\right.
\label{eq:Poisson-particular-gated}
\end{eqnarray}
In this case the 2D charge density will be the same as that in Eq.\,\eqref{eq:rho-2D} [except $q_u$ should be substituted by $q_g$]. Substitution of the 2D charge density and the particular solution of Poisson equation \eqref{eq:Poisson-particular-gated} into Eq.\,\eqref{eq:Poisson-integrated} results in the dispersion relation for gated plasmons
\begin{equation}
\epsilon_{1}+\frac{\epsilon_{2}}{{\rm tanh}(q_gd)}=\frac{q_a\sigma(\omega)}{i\omega\epsilon_{0}}.
\end{equation}
For small distances between graphene and metal, $dq<<1$ , ${\rm tanh}(q_gd)\simeq q_gd$ and $\epsilon_1\ll\epsilon_2/(q_gd)$. In this case the dispersion relation for gated plasmons can be expressed as
\begin{equation}
q_{g}=\omega\sqrt{\frac{\hbar\epsilon_{2}}{4E_{F}c\alpha d}}.
\end{equation}

\section{ Derivation of energy-dependent hopping amplitudes}

\subsection{Transfer matrix}

We begin by determining the general transfer matrix relating, in one
specific region labeled by $j=u,g$, the functions at $x$ and $x+\Delta x$:
\begin{equation}
\left(\begin{array}{c}
\phi_{j}(x+\Delta x)\\
\sigma_{j}\phi'_{j}(x+\Delta x)
\end{array}\right)=\mathbf{W}_{j}(\Delta x)\left(\begin{array}{c}
\phi_{j}(x)\\
\sigma_{j}\phi'_{j}(x)
\end{array}\right)
\end{equation}
where $\phi'_{j}(x)=\frac{\partial\phi_{j}(x)}{\partial x}$ and $\mathbf{W}_{j}(\Delta x)$
is the corresponding transfer matrix in the region labeled by $j$.
The transfer-matrix coefficients can be found by expanding the expression
above, using the definitions of $\phi(x)$ in Eqs.~ (\ref{eq:phi_u and phi_g_1}) and 
(\ref{eq:phi_u and phi_g_2}) 

\begin{widetext}

\begin{equation}
\begin{array}{c}
\phi_{j}^{+}e^{iq_{j}x}e^{iq_{j}\Delta x}+\phi_{j}^{-}e^{-iq_{j}x}e^{-iq_{j}\Delta x}=\mathbf{W}_{j}(\Delta x)_{11}\left(\phi_{j}^{+}e^{iq_{j}x}+\phi_{j}^{-}e^{-iq_{j}x}\right)+\mathbf{W}_{j}(\Delta x)_{12}i\sigma_{j}q_{j}\left(\phi_{j}^{+}e^{iq_{j}x}-\phi_{j}^{-}e^{-iq_{j}x}\right)\\
\\
i\sigma_{j}q_{j}\left(\phi_{j}^{+}e^{iq_{j}x}e^{iq_{j}\Delta x}-\phi_{j}^{-}e^{-iq_{j}x}e^{-iq_{j}\Delta x}\right)=\mathbf{W}_{j}(\Delta x)_{21}\left(\phi_{j}^{+}e^{iq_{j}x}+\phi_{j}^{-}e^{-iq_{j}x}\right)+\mathbf{W}_{j}(\Delta x)_{22}i\sigma_{j}q_{j}\left(\phi_{j}^{+}e^{iq_{j}x}-\phi_{j}^{-}e^{-iq_{j}x}\right)
\end{array}
\end{equation}
Rearranging terms related to the same coefficients:
\begin{equation}
\begin{array}{c}
\phi_{j}^{+}e^{iq_{j}x}e^{iq_{j}\Delta x}+\phi_{j}^{-}e^{-iq_{j}x}e^{-iq_{j}\Delta x}=\phi_{j}^{+}e^{iq_{j}x}\left(\mathbf{W}_{j}(\Delta x)_{11}+\mathbf{W}_{j}(\Delta x)_{12}i\sigma_{j}q_{j}\right)+\phi_{j}^{-}e^{-iq_{j}x}\left(\mathbf{W}_{j}(\Delta x)_{11}-\mathbf{W}_{j}(\Delta x)_{12}i\sigma_{j}q_{j}\right)\\
\\
\phi_{j}^{+}e^{iq_{j}x}e^{iq_{j}\Delta x}-\phi_{j}^{-}e^{-iq_{j}x}e^{-iq_{j}\Delta x}=\phi_{j}^{+}e^{iq_{j}x}\left(\frac{\mathbf{W}_{j}(\Delta x)_{21}+\mathbf{W}_{j}(\Delta x)_{22}i\sigma_{j}q_{j}}{i\sigma_{j}q_{j}}\right)-\phi_{j}^{-}e^{-iq_{j}x}\left(\frac{\mathbf{W}_{j}(\Delta x)_{22}i\sigma_{j}q_{j}-\mathbf{W}_{j}(\Delta x)_{21}}{i\sigma_{j}q_{j}}\right)
\end{array}
\end{equation}
The exponentials of $\Delta x$ must then obey:
\begin{equation}
\begin{array}{c}
e^{iq_{j}\Delta x}=\mathbf{W}_{j}(\Delta x)_{11}+\mathbf{W}_{j}(\Delta x)_{12}i\sigma_{j}q_{j}=\frac{\mathbf{W}_{j}(\Delta x)_{21}+\mathbf{W}_{j}(\Delta x)_{22}i\sigma_{j}q_{j}}{i\sigma_{j}q_{j}}\\
\\
e^{-iq_{j}\Delta x}=\mathbf{W}_{j}(\Delta x)_{11}-\mathbf{W}_{j}(\Delta x)_{12}i\sigma_{j}q_{j}=\frac{\mathbf{W}_{j}(\Delta x)_{22}i\sigma_{j}q_{j}-\mathbf{W}_{j}(\Delta x)_{21}}{i\sigma_{j}q_{j}}
\end{array}
\end{equation}
Expanding the exponentials into cosine and sine and solving the system
of equations, it follows that for arbitrary $\Delta x$ the transfer-matrix coefficients are:
\begin{equation}
\begin{array}{cc}
\mathbf{W}_{j}(\Delta x)_{11}={\rm cos}(q_{j}\Delta x), & \mathbf{W}_{j}(\Delta x)_{12}=\frac{{\rm sin}(q_{j}\Delta x)}{\sigma_{j}q_{j}}\\
\mathbf{W}_{j}(\Delta x)_{21}{\rm =-\sigma_{j}q_{j}sin}(q_{j}\Delta x), & \mathbf{W}_{j}(\Delta x)_{22}={\rm cos}(q_{j}\Delta x)
\end{array}\label{eq:transfer_matrix_W}
\end{equation}

\subsection{$\sigma_{g}\phi'_{g}(x)$ as a function of $\phi_{g}(nl),\phi_{g}(nl+l_{g})$:}

For the gated region in the $n^{th}$ unit cell we define: 
\begin{equation}
\left(\begin{array}{c}
\phi_{g}(nl+l_{g})\\
\sigma_{g}\phi'_{g}(nl+l_{g})
\end{array}\right)=\mathbf{W}_{g}(l_{g})\left(\begin{array}{c}
\phi_{g}(nl)\\
\sigma_{g}\phi'_{g}(nl)
\end{array}\right)\label{eq:W_g(l_g)}
\end{equation}
\begin{equation}
\left(\begin{array}{c}
\phi_{g}(x)\\
\sigma_{g}\phi'_{g}(x)
\end{array}\right)=\mathbf{W}_{g}(x-nl)\left(\begin{array}{c}
\phi_{g}(nl)\\
\sigma_{g}\phi'_{g}(nl)
\end{array}\right)\label{eq:W_g(x-nl)}
\end{equation}
Using the first row of the transfer matrix, Eq.~(\ref{eq:W_g(l_g)})
leads to
\begin{equation}
\phi_{g}(nl+l_{g})={\rm cos}(q_{g}l_{g})\phi_{g}(nl)+\frac{{\rm sin}(q_{g}l_{g})}{\sigma_{g}q_{g}}\sigma_{g}\phi'_{g}(nl)
\end{equation}
which can be rearranged as
\begin{equation}
\sigma_{g}\phi'_{g}(nl)=\frac{\sigma_{g}q_{g}\phi_{g}(nl+l_{g})}{{\rm sin}(q_{g}l_{g})}-\frac{{\rm \sigma_{g}q_{g}cos}(q_{g}l_{g})\phi_{g}(nl)}{{\rm sin}(q_{g}l_{g})}.\label{eq:sigma_g phi'_g(nl)}
\end{equation}
Using the second row of Eq.~(\ref{eq:W_g(x-nl)}) leads to:
\begin{equation}
\sigma_{g}\phi'_{g}(x)={\rm -\sigma_{g}q_{g}sin}(q_{g}\left[x-nl\right])\phi_{g}(nl)+{\rm cos}(q_{g}\left[x-nl\right])\sigma_{g}\phi'_{g}(nl)
\end{equation}
which can be substituted into Eq.~(\ref{eq:sigma_g phi'_g(nl)}). After some manipulation and using trigonometric identities, this yields:
\begin{equation}
\sigma_{g}\phi'_{g}(x)=-\phi_{g}(nl)\frac{\sigma_{g}q_{g}{\rm cos}(q_{g}\left[l_{g}+nl-x\right])}{{\rm sin}(q_{g}l_{g})}+\phi_{g}(nl+l_{g})\frac{\sigma_{g}q_{g}{\rm cos}(q_{g}\left[x-nl\right])}{{\rm sin}(q_{g}l_{g})}.\label{eq:sigma_g phi'_g(x)}
\end{equation}

\subsection{$\sigma_{u}\phi'_{u}(x)$ as a function of $\phi_{u}(nl+l),\phi_{u}(nl+l_{g})$:}

For the ungated region in the $n$th unit cell we define: 
\begin{equation}
\left(\begin{array}{c}
\phi_{u}(nl+l)\\
\sigma_{u}\phi'_{u}(nl+l)
\end{array}\right)={\bf W}_{u}(l_{u})\left(\begin{array}{c}
\phi_{u}(nl+l_{g})\\
\sigma_{u}\phi'_{u}(nl+l_{g})
\end{array}\right),
\end{equation}
\begin{equation}
\left(\begin{array}{c}
\phi_{u}(x)\\
\sigma_{u}\phi'_{u}(x)
\end{array}\right)={\bf W}_{u}(x-nl-l_{g})\left(\begin{array}{c}
\phi_{u}(nl+l_{g})\\
\sigma_{u}\phi'_{u}(nl+l_{g})
\end{array}\right).
\end{equation}
This situation is exactly like the one for the gated region, but with
the changes 
\[
\begin{array}{rcl}
{\rm cos}(q_{g}l_{g}),{\rm sin}(q_{g}l_{g})&\longrightarrow&{\rm cos}(q_{u}l_{u}),{\rm sin}(q_{u}l_{u})\\
\phi_{g}(nl)&\longrightarrow&\phi_{u}(nl+l_{g})\\
\phi_{g}(nl+l_{g})&\longrightarrow&\phi_{u}(nl+l)\\
{\bf W}_{g}(x-nl)&\longrightarrow&{\bf W}_{u}(x-nl-l_{g})
\end{array}
\]
With these substitutions, we find
\begin{equation}
\sigma_{u}\phi'_{u}(x)=-\phi_{u}(nl+l_{g})\frac{\sigma_{u}q_{u}{\rm cos}(q_{u}\left[nl+l-x\right])}{{\rm sin}(q_{u}l_{u})}+\phi_{u}(nl+l)\frac{\sigma_{u}q_{u}{\rm cos}(q_{u}\left[x-nl-l_{g}\right])}{{\rm sin}(q_{u}l_{u})}.\label{eq:eq:sigma_u phi'_u(x)}
\end{equation}

\subsection{The tight-binding equation}

With Eqs.~(\ref{eq:sigma_g phi'_g(x)}) and (\ref{eq:eq:sigma_u phi'_u(x)})
we managed to write, for a given region, a function describing the
plasmon in an arbitrary position $x$ in terms of functions at the
boundaries. Now, we equate both Eqs.~(\ref{eq:sigma_g phi'_g(x)}) and
(\ref{eq:eq:sigma_u phi'_u(x)}) at $x=nl$ and $x=nl+l_{g}$ to obtain
the tight-binding equations (remember that the indexes $u$ and $g$ are
irrelevant on the wavefunctions now):
At $x=nl$ (note that at this point, $\sigma_{g}\phi'_{g}(x)$ is
evaluated in the $n$th unit cell while $\sigma_{u}\phi'_{u}(x)$
is evaluated at the $(n-1)$th unit cell):
\begin{equation}
\sigma_{g}\phi'_{g}(nl)=\sigma_{u}\phi'_{u}([n-1]l)
\end{equation}
\begin{equation}
-\phi(nl)\frac{\sigma_{g}q_{g}{\rm cos}(q_{g}l_{g})}{{\rm sin}(q_{g}l_{g})}+\phi(nl+l_{g})\frac{\sigma_{g}q_{g}}{{\rm sin}(q_{g}l_{g})}=-\phi_{u}([n-1]l+l_{g})\frac{\sigma_{u}q_{u}}{{\rm sin}(q_{u}l_{u})}+\phi_{u}([n-1]l+l)\frac{\sigma_{u}q_{u}{\rm cos}(q_{u}\left[x-(n-1)l-l_{g}\right])}{{\rm sin}(q_{u}l_{u})}
\end{equation}
\begin{equation}
\phi(nl+l_{g})\frac{\sigma_{g}q_{g}}{{\rm sin}(q_{g}l_{g})}+\phi(nl-lu)\frac{\sigma_{u}q_{u}}{{\rm sin}(q_{u}l_{u})}=\phi(nl)\left[\frac{\sigma_{u}q_{u}{\rm cos}(q_{u}l_{u})}{{\rm sin}(q_{u}l_{u})}+\frac{\sigma_{g}q_{g}{\rm cos}(q_{g}l_{g})}{{\rm sin}(q_{g}l_{g})}\right]\label{eq:TB_eq_12}
\end{equation}
At $x=nl+l_{g}$:
\begin{equation}
\sigma_{g}\phi'_{g}(nl+l_{g})=\sigma_{u}\phi'_{u}(nl+l_{g})
\end{equation}
\begin{equation}
-\phi(nl)\frac{\sigma_{g}q_{g}}{{\rm sin}(q_{g}l_{g})}+\phi(nl+l_{g})\frac{\sigma_{g}q_{g}{\rm cos}(q_{g}l_{g})}{{\rm sin}(q_{g}l_{g})}=-\phi(nl+l_{g})\frac{\sigma_{u}q_{u}{\rm cos}(q_{u}l_{u})}{{\rm sin}(q_{u}l_{u})}+\phi(nl+l)\frac{\sigma_{u}q_{u}}{{\rm sin}(q_{u}l_{u})}
\end{equation}

\noindent where we used the first representation of $\sigma_{u}\phi'_{u}(x)$.
\begin{equation}
\phi(nl)\frac{\sigma_{g}q_{g}}{{\rm sin}(q_{g}l_{g})}+\phi(nl+l)\frac{\sigma_{u}q_{u}}{{\rm sin}(q_{u}l_{u})}=\phi(nl+l_{g})\left[\frac{\sigma_{u}q_{u}{\rm cos}(q_{u}l_{u})}{{\rm sin}(q_{u}l_{u})}+\frac{\sigma_{g}q_{g}{\rm cos}(q_{g}l_{g})}{{\rm sin}(q_{g}l_{g})}\right]\label{eq:TB_eq_22}
\end{equation}
With Eqs.~(\ref{eq:TB_eq_12}) and (\ref{eq:TB_eq_22}) we have the two
desired tight-binding equations.

\end{widetext} 


%

\end{document}